\documentclass[aps,prd,eqsecnum,showpacs,floatfix]{revtex4}
\usepackage{color,graphicx}
\usepackage{subfigure}
\usepackage{amsfonts}
\usepackage{amssymb}
\begin{document}


\title{
Einstein-Rosen waves and the
self-similarity hypothesis in cylindrical symmetry}

\author{
$^{1}$Tomohiro Harada\footnote{harada@rikkyo.ac.jp},
$^{2}$ Ken-ichi Nakao\footnote{knakao@sci.osaka-cu.ac.jp} and
$^{3}$ Brien C. Nolan\footnote{brien.nolan@dcu.ie}
}
\affiliation{
$^{1}$Department of Physics, Rikkyo University, Toshima, Tokyo 171-8501,
Japan\\
$^{2}$Department of Physics, Osaka City University, Osaka 558-8585, Japan\\
$^{3}$School of Mathematical Sciences, Dublin City University, Glasnevin, Dublin 9, Ireland
}
\date{\today}

\begin{abstract}
The self-similarity hypothesis claims that in classical general
 relativity, spherically symmetric solutions may naturally evolve to a
 self-similar form in certain circumstances. In this context, the
 validity of the corresponding hypothesis in nonspherical geometry is
 very interesting as there may exist
 gravitational waves. We investigate self-similar vacuum solutions to
 the Einstein equation in the so-called whole-cylinder symmetry.
 We find that those
 solutions are reduced to part of the Minkowski spacetime with a regular or
 conically singular axis and with trivial or nontrivial topology if the
 homothetic vector is orthogonal to the cylinders of symmetry.
 These solutions are analogous to the Milne universe, but
 only in the direction parallel to the axis.
 Using these solutions, we
 discuss the nonuniqueness (and nonvanishing nature) of $C$ energy and the existence of a
cylindrical trapping horizon in Minkowski spacetime. Then, as we
generalize
 the analysis, we find a two-parameter family of self-similar vacuum
 solutions, where the
 homothetic vector is not orthogonal to the 
cylinders in general.
The family includes the Minkowski, the Kasner and the
cylindrical Milne solutions.
The obtained solutions
 describe
the interior to the exploding (imploding) shell of
gravitational waves or the collapse (explosion)
of gravitational waves involving singularities
from nonsingular initial data in general.
Since recent
 numerical simulations strongly suggest that one of these solutions
may describe the asymptotic behavior of gravitational waves from
the collapse of a dust cylinder, this means that the self-similarity
hypothesis is naturally generalized to cylindrical symmetry.

\end{abstract}

\maketitle


\section{Introduction}

In studying nonspherical vacuum gravitational fields, cylindrically
symmetric systems have the advantages of being essentially 1+1
dimensional, and, unlike the spherical case, possessing a dynamical
degree of freedom in gravity, i.e., gravitational waves. Solutions
in this system are discussed in~\cite{Einstein:1937} and are called
Einstein-Rosen waves. These have played an important part in the
history of gravitational wave research, principally in elucidating
the reality of gravitational waves as carriers of
energy~\cite{Kennefick:2007}. Thus, several researchers have studied
vacuum and nonvacuum cylindrical systems in an attempt to clarify
the nature of (especially) nonspherical gravitational collapse, in
a system that involves the essential nonlinearity of the
gravitational field and the emission of gravitational waves but
requires the analysis of partial differential equations with just
one spatial dimension. See for
example~\cite{Thorne:1965,Melvin:1965,Apos:1992,echeverria:1993,letelier:1994,Chiba:1996,Ashtekar:1997,Hayward:2000,Nolan:2002,Wang:2003,Nakao:2004,
Nakao:2005,Kurita:2006,Nakao:2007,Nakao:2008}.

As in many other fundamental theories, self-similarity plays an
important role in gravitation. This importance is encapsulated in
Carr's self-similarity hypothesis~\cite{Carr:1993,Carr:2005}, which
originally asserts that in
the cosmological context, spherically symmetric
solutions of Einstein's equations evolve to a
self-similar form. This evolution can be towards either an
intermediate or an endstate.
Later it was found that this is actually the
case in gravitational collapse.
Examples include the collapse of a
soft fluid sphere~\cite{Harada:2001,Snajdr:2006} and
critical phenomena emerging in
the spherical collapse of a variety of matter
fields~\cite{Gundlach:2007}.
We should also note that
spatially homogeneous cosmological models
provide an example of tendency towards self-similar
solutions in non spherically symmetric systems: see~\cite{Wainwright:1997} for a review.

Spherically symmetric self-similar spacetimes have been extensively
studied and are now well understood. A natural next step therefore is to
consider the role of self-similarity in cylindrical systems. These arise as a special case of $G2$ spacetimes -- i.e.\ spacetimes admitting a two-dimensional group of isometries, usually but not always Abelian. There have been several studies of self-similar $G2$ \textit{cosmological} models -- again see~\cite{Wainwright:1997}, but there has been little work heretofore on the role of self-similarity in cylindrical collapse.
Indeed it is generally true that in the cylindrical case,
results are fewer and farther between: the additional degrees of
freedom that exist in the cylindrical case, which render the
systems of equations encountered considerably more difficult, have
not to date allowed the development of as clear a picture as we have
in the spherical case (but
see~\cite{Wang:2003,Sharif:2005a,Sharif:2005b,Nolan:2007}). We note that
Ref.~\cite{Nolan:2007}
deals with self-similar cylindrical dust collapse.

The present work seeks to increase our understanding of the role of
self-similarity in cylindrical gravitational collapse, and the
veracity of the similarity hypothesis in this context. In
particular, this work is motivated by numerical results obtained by
two of the present authors (KN and TH) and their collaborators which show that
self-similar cylindrically symmetric vacuum solutions appear to
describe the asymptotic behavior of the (numerically obtained)
gravitational field outside a collapsing hollow dust
cylinder~\cite{Nakao:2009}.
We present here a two-parameter family of self-similar {\it vacuum}
solutions in cylindrical symmetry.

This paper is organized as follows. In Sec. II, we present
formulations for vacuum spacetimes and self-similar spacetimes in
the so-called whole-cylinder symmetry
independently. In Sec. III, we solve the
Einstein field equations for vacuum self-similar spacetimes and
obtain exact solutions. The motivation for this is to twofold. First, we wish to study self-similar cylindrical collapse of pure gravitational waves, as representing the simplest example of self-similar cylindrical collapse. The hope is that in some sense, this would provide the standard model for self-similar cylindrical collapse in the same way that the Schwarzschild solution provides the standard model of the (endstate of) spherically symmetric collapse. The second motivation is to determine the description of possible exteriors of collapsing self-similar, cylindrical matter. We recall also the overarching motivation for this study: to determine self-similar solutions that may act as (intermediate) attractors of more general cylindrically symmetric spacetimes. In fact we show that Einstein-Rosen waves with
a homothetic vector orthogonal to the cylinders of symmetry are \textit{flat} and reduce to
part of the
Minkowski spacetime with or without a conical singularity and
with or without nontrivial topology.
Moreover, we show that the above metric form
of part of the Minkowski spacetime implies nontrivial (i.e.\ nonzero)
$C$ energy~\cite{Thorne:1965} and a trapping horizon. In Sec. IV, we naturally extend
the solutions to more general class and find that these solutions
are also self-similar where the homothetic vector is not orthogonal to the 
cylinders of symmetry
in general. We consider the analytical extension of the
solutions, analyze the global structure of the obtained spacetimes
and find that these describe interesting nonlinear dynamics of
gravitational waves.
We discuss the physical interpretation of these solutions.
In Sec. V, we summarize the paper. We use the units, in which $G=c=1$.

\section{Cylindrical symmetry and self-similarity}
\label{sec:basic}

\subsection{Spacetimes in whole-cylinder symmetry}

For cylindrically symmetric spacetimes, we assume that
there are two commuting, spatial Killing vectors $(\xi_{(1)},\xi_{(2)})$
such that the
orthogonal space is integrable and the Killing coordinate
$\varphi$, where $\xi_{(1)}=\partial/\partial \varphi$,
is identified at $0$ and
$2\pi$.
We call $\xi_{(1)}$ and $\xi_{(2)}$ azimuthal and translational
Killing vectors, respectively.
Here we shall additionally assume that each of the two Killing
vectors be hypersurface orthogonal, which is called
whole-cylinder symmetry~\cite{Thorne:1965,Melvin:1965}
or the polarized case~\cite{Hayward:2000}.
The circumferential radius $\rho$, the specific length $\ell$
and the areal radius $r$
are defined as
\begin{equation}
\rho^{2}:=\xi_{(1) a}\xi_{(1)}^{a},\quad
\ell^{2}:=\xi_{(2) a}\xi_{(2)}^{a}\quad \mbox{and}\quad
r:=\rho \ell.
\end{equation}
Note then that $r$ is the areal radius of the orbits of the isometry group and so $r\geq 0$ with $r=0$ at the axis.
The line element in this class of spacetimes
is given by~\cite{Stephani:2003}:
\begin{equation}
ds^{2}=-2e^{2\gamma(u,v)-2\psi(u,v)}dudv+e^{-2\psi(u,v)}r^{2}(u,v)d\varphi^{2}
+e^{2\psi(u,v)}dz^{2}.
\label{eq:metric_cylindrical}
\end{equation}
We note that this form of the line element is unchanged under rescalings of the null coordinates $u\to\bar{u}(u)$, $v\to\bar{v}(v)$. The regular axis condition~\cite{Stephani:2003,Hayward:2000} ensures the
ratio of an infinitesimal circle around the
axis to its diameter to be $\pi$.

The $C$ energy $E$ is then defined
as~\cite{Thorne:1965,Chiba:1996,Hayward:2000}
\begin{eqnarray}
E:=\frac{1}{8}\left(1-\ell^{-2}\nabla^{a}r\nabla _{a} r\right).
\end{eqnarray}
This quantity is assumed to represent the line
energy density enclosed inside the cylinder. A cylinder, which is a two-surface
given by $u=$const and $v=$const, is said to be trapped, marginally
trapped and untrapped if $\nabla^{a}r$ is timelike, null and
spacelike, respectively. In terms of the $C$ energy, a cylinder is
trapped, marginally trapped and untrapped if $E>1/8$, $E=1/8$ and
$E<1/8$, respectively. A cylindrical trapping horizon is a
hypersurface foliated by marginal cylinders.

The Einstein equations $G_{\mu\nu}=8\pi T_{\mu\nu}$ reduce to
\begin{eqnarray}
r_{,uu}+2r\psi_{,u}^{2}-2\gamma_{,u}r_{,u}&=&8\pi r T_{uu},
\label{eq:Einstein_uu}\\
r_{,vv}+2r\psi_{,v}^{2}-2\gamma_{,v}r_{,v}&=&-8\pi r T_{vv},
\label{eq:Einstein_vv}\\
r_{,uv}&=&8\pi r T_{uv},
\label{eq:Einstein_uv}\\
-2r^{2}e^{-2\gamma}(\gamma_{,uv}+\psi_{,u}\psi_{,v})
&=&8\pi T_{\varphi\varphi},
\label{eq:Einstein_pp}\\
e^{4\psi-2\gamma}(2r \psi_{,uv}+r_{,u}\psi_{,v}+r_{,v}\psi_{,u}-
r\gamma_{,uv}-r_{,uv}-r\psi_{,u}\psi_{,v})&=& 4\pi r T_{zz},
\label{eq:Einstein_zz}
\end{eqnarray}
where $T_{\mu\nu}$ is the stress-energy tensor for the matter fields.

\subsection{Einstein-Rosen waves}
Here we consider vacuum spacetimes in whole-cylinder symmetry.
Solutions of the field equations in this case are called
Einstein-Rosen waves~\cite{Einstein:1937}. For such a
cylindrically symmetric vacuum spacetime, in which the line element is given by
Eq.~(\ref{eq:metric_cylindrical}), Eq.~(\ref{eq:Einstein_uv})
reduces to
\begin{equation}
r_{,uv}=0,
\end{equation}
implying
\begin{equation}
r=f(u)+g(v),
\label{eq:f_g}
\end{equation}
where $f$ and $g$ are arbitrary functions.

If $\nabla^{a}r$ is spacelike, we can choose $f$ and $g$
by rescaling $u$ and $v$ such that
\begin{equation}
r=\frac{v-u}{\sqrt{2}}.
\label{eq:r_v-u}
\end{equation}
Introducing the time and radial coordinates
\begin{equation}
t=\frac{v+u}{\sqrt{2}}\quad \mbox{and}\quad x=\frac{v-u}{\sqrt{2}},
\end{equation}
we obtain the metric in the form
\begin{equation}
ds^{2}=
e^{2(\gamma-\psi)}(-dt^{2}+dx ^{2})
+e^{-2\psi}x^{2} d\varphi^{2}
+e^{2\psi}dz^{2}.
\label{eq:metric_t_x_x}
\end{equation}
We note that this line element, subject to the Einstein equations below, corresponds to the original Einstein-Rosen waves. That is, the Einstein-Rosen paper \cite{Einstein:1937} deals exclusively with the case where $\nabla^ar$ is spacelike. For convenience, we will use the term to refer to any vacuum whole-cylinder symmetric solution of the Einstein equations. 

The nontrivial components of the Einstein equations
become the following simple set of partial differential equations:
\begin{eqnarray}
&& -\psi_{,tt}+\psi_{,xx}+\frac{1}{x}\psi_{,x}=0,
\label{eq:wave_spacelike}\\
&& \gamma_{,x}=x(\psi_{,x}^{2}+\psi_{,t}^{2}),
\label{eq:gammax_spacelike}\\
&& \gamma_{,t}=2x \psi_{,x}\psi_{,t}.
\label{eq:gammat_spacelike}
\end{eqnarray}
Equation~(\ref{eq:Einstein_pp}) is automatically satisfied due to
Eqs.~(\ref{eq:gammax_spacelike}) and (\ref{eq:gammat_spacelike}).
The regular axis condition reduces to $\gamma\to 0$ as $x\to 0$
in this coordinate system.

If $\nabla^{a}r$ is timelike, we can choose $f$ and $g$
by rescaling $u$ and $v$ such that
\begin{equation}
r=\frac{v+u}{\sqrt{2}}=t.
\label{eq:r_v+u}
\end{equation}
Then, we get the metric in the form
\begin{equation}
ds^{2}=
e^{2(\gamma-\psi)}(-dt^{2}+dx ^{2})
+e^{-2\psi}t^{2} d\varphi^{2}
+e^{2\psi} dz^{2}.
\label{eq:metric_t_x_t}
\end{equation}
The nontrivial components of the Einstein equations
become the following simple set of ordinary differential equations:
\begin{eqnarray}
&& -\psi_{,tt}+\psi_{,xx}-\frac{1}{t}\psi_{,t}=0, \\
&& \gamma_{,t}=t(\psi_{,t}^{2}+\psi_{,x}^{2}), \\
&& \gamma_{,x}=2t \psi_{,t}\psi_{,x}.
\end{eqnarray}
We can see that the equations are
the same as those for spacelike $\nabla^{a}r$ if we exchange $t$ and $x$.

If $\nabla^{a}r$ is null, we can choose $f$ and $g$ such that
\begin{equation}
r=u.
\end{equation}
The Einstein equations reduce to
\begin{eqnarray}
&& \psi_{,uv}=0, \\
&& \psi_{,v}=0, \\
&& \gamma_{,u}=u\psi_{,u}^{2}.
\end{eqnarray}
$\psi=\psi(u)$ follows from the above.

\subsection{Self-similar spacetimes in whole-cylinder symmetry}

 We now
consider the case where the spacetime is self-similar as well as
cylindrically symmetric: there is no {\em a priori} reason to
suppose that this will lead to trivial solutions only. In other
words, we assume that the spacetime admits a vector field ${\bf v}$
which satisfies the following equation
\begin{equation}
{\cal L}_{{\bf v}}g_{\mu\nu}=2g_{\mu\nu},
\label{eq:homothetic_Killing_eq}
\end{equation}
where ${\cal L}_{{\bf v}}$ denotes the Lie derivative along ${\bf v}$.
We refer to the vector field ${\bf v}$ and
Eq.~(\ref{eq:homothetic_Killing_eq})
as the homothetic vector and the homothetic equation,
respectively.

We assume that ${\bf v}$ has the following form:
\begin{equation}
{\bf v}=\alpha(u,v)\frac{\partial}{\partial u}+\beta(u,v)
\frac{\partial}{\partial v},
\end{equation}
where ${\bf v}$ is then assumed to be orthogonal to the cylinders of symmetry. We will refer to this as a cylindrical homothetic vector.
We will mention the limitation of this assumption later.
Then, the homothetic equations~(\ref{eq:homothetic_Killing_eq}) yield
$\alpha=\alpha(u)$ and $\beta=\beta (v)$.
We can then generically make the coordinate transformation
$\bar{u}=\bar{u}(u)$ and $\bar{v}=\bar{v}(v)$ satisfying
the following relations:
\begin{equation}
\alpha(u)\frac{d\bar{u}}{d u}=2 \bar{u}
\quad
\mbox{and}
\quad
\beta (v)\frac{d\bar{v}}{dv}=2\bar{v}.
\end{equation}
Then, we can have
\begin{equation}
{\bf v}=2u \frac{\partial}{\partial u}+2v
\frac{\partial}{\partial v},
\label{eq:standard_HKV}
\end{equation}
where here and hereafter we omit bars for simplicity. The homothetic
vector is timelike, spacelike and null if $uv$ is positive, negative
and zero, respectively. If $\alpha=0$ or $\beta=0$, the homothetic
vector is null and we do not consider these special cases here.

It is then straightforward to show that the homothetic equations~(\ref{eq:homothetic_Killing_eq}) lead to
\begin{equation}
 e^{2\psi}=|u| e^{2 P(\eta)}, \quad r = |u| S(\eta),\quad e^{2\gamma}=e^{2G (\eta)}
 \label{eq:metricfns_ss_uv}\end{equation}
 where $\eta=v/u$ and $P,S$ and $G$ are arbitrary functions.

Thus, we obtain the
following standard form of the metric for whole-cylindrically
symmetric self-similar spacetimes:
\begin{equation}
ds^{2}=-2e^{2G(\eta)-2P(\eta)}|u|^{-1}dudv+e^{-2P(\eta)}|u|
S^{2}(\eta) d\varphi^{2} +e^{2P(\eta)}|u| dz^{2}.
\label{eq:metric_ss_uv}
\end{equation}

We can substitute the above form into the Einstein equations
(\ref{eq:Einstein_uu}) -- (\ref{eq:Einstein_zz}) and obtain a set of
ordinary differential equations for $P$, $G$ and $S$.
Fortunately, for the vacuum
case, we can greatly simplify the system.

\section{Einstein-Rosen waves with a cylindrical homothetic vector}
\subsection{One-parameter family of solutions}
We consider self-similar vacuum solutions in this section.
It is not trivial
that the choice of $f$ and $g$ adopted in Sec.~\ref{sec:basic}B
is compatible with the self-similarity introduced
in Sec.~\ref{sec:basic}C. In fact, 
for $r$ to be compatible with the self-similarity, i.e.,
Eqs.~(\ref{eq:metricfns_ss_uv}),
$f$ and $g$ introduced in Eq.~(\ref{eq:f_g})
for the vacuum solution must satisfy
\begin{equation}
f=C_{1}u, \quad \mbox{and} \quad g=C_{2}v,
\end{equation}
where $C_{1}$ and $C_{2}$ are arbitrary constants.
(We note that a trivial addition of a constant may be required to obtain the above form.)
Hence, the choices of $f$ and $g$ given by
Eqs.~(\ref{eq:r_v-u}) and (\ref{eq:r_v+u}) are both
compatible with self-similarity.

For the moment we restrict ourselves to the case where
$r$ has a spacelike gradient.
Now we can adopt $(t, x)$ coordinates and by combining Eqs.~(\ref{eq:metric_t_x_x}), (\ref{eq:metricfns_ss_uv}) and (\ref{eq:metric_ss_uv}), we find the metric
in the following form:
\begin{equation}
ds^{2}=
e^{2\Gamma(\xi)-2\Psi(\xi)}x^{-1}(-dt^{2}+dx ^{2})
+e^{-2\Psi(\xi)}x d\varphi^{2}
+e^{2\Psi(\xi)}x dz^{2},
\end{equation}
where
\begin{equation}
\xi=\frac{t}{x}.
\end{equation}
(We recall that $x=r\geq0$.)
In this case, the Einstein equations reduce to a set of simple
ordinary differential equations. Noting
\begin{eqnarray}
\psi&=&\Psi(\xi)+\frac{1}{2}\ln x,
\label{eq:selfsimilar_psi_spacelike}\\
\gamma&=&\Gamma(\xi),
\label{eq:selfsimilar_gamma_spacelike}
\end{eqnarray}
Eqs.~(\ref{eq:wave_spacelike}), (\ref{eq:gammax_spacelike})
and (\ref{eq:gammat_spacelike}) reduce
to the following ordinary differential equations:
\begin{eqnarray}
&& (\xi^{2}-1)\Psi''+\xi \Psi'=0,
\label{eq:selfsimilar_wave}\\
&& \xi \Gamma'=-\left(\xi \Psi'-\frac{1}{2}\right)^{2}-\Psi'^{2},
\label{eq:Gamma1}\\
&& \Gamma'=-2 \left(\xi \Psi'-\frac{1}{2}\right)\Psi',
\label{eq:Gamma2}
\end{eqnarray}
where the prime denotes the derivative with respect to $\xi$. From 
Eqs.~(\ref{eq:Gamma1}) and (\ref{eq:Gamma2}), we obtain
\begin{equation}
(\xi^{2}-1)\Psi'^{2}-\frac{1}{4}=0.
\end{equation}
Therefore, $\xi^{2}>1$ and
\begin{equation}
\Psi'=\pm \frac{1}{2\sqrt{\xi^{2}-1}}.
\label{eq:Phi'}
\end{equation}
This satisfies Eq.~(\ref{eq:selfsimilar_wave}).
We can integrate the above equation and obtain
\begin{equation}
\Psi= \frac{1}{2}\ln \left|\xi\pm \sqrt{\xi^{2}-1}
\right| +\Psi_{0},
\label{eq:selfsimilar_Psi_spacelike}
\end{equation}
where $\Psi_{0}$ is a constant of integration.
$\Gamma$ is obtained by substituting Eq.~(\ref{eq:Phi'}) into
Eq.~(\ref{eq:Gamma2}) and integrating the resultant equation.
The result is
\begin{equation}
\Gamma=\frac{1}{2}\ln \left|\frac{1}{2}
\left(
\frac{\xi}{\sqrt{\xi^{2}-1}}\pm 1\right)\right|
+\lambda,
\label{eq:Gamma_spacelike_r}
\end{equation}
where $\lambda$ is a constant of integration.
Getting back to the original metric functions $\psi$ and $\gamma$,
the solution is given by
\begin{eqnarray}
\psi&=& \frac{1}{2}\ln |\xi\pm \sqrt{\xi^{2}-1}| + \Psi_{0}+\frac{1}{2}\ln x,
\\
\gamma&=&\frac{1}{2}\ln \left|\frac{1}{2}\left(
\frac{\xi}{\sqrt{\xi^{2}-1}}\pm 1\right)\right|
+\lambda.
\end{eqnarray}
We can assume $t>0$ because the flip of the sign of $t$ corresponds to
the other branch of solutions.
Note that we can set $\Psi_{0}=0$ by absorbing it into the coordinates
as follows:
\begin{equation}
\tilde{t}=e^{-2\Psi_{0}} t, \tilde{x}=x e^{-2\Psi_{0}}, \tilde{\varphi}=\varphi,
\tilde{z}=e^{2\Psi_{0}}z.
\end{equation}
Up to this gauge parameter,
the solutions are parametrized by $\lambda$.

For the timelike $\nabla^{a}r$ case, we can
obtain the solution just by exchanging $t$ and $x$
in the solution for the spacelike $\nabla^{a}r$ case.
This corresponds to the replacement of $\xi$ with $\xi^{-1}$.
The solution is therefore given by
\begin{eqnarray}
\psi&=& \frac{1}{2}\ln |\xi^{-1} \pm
\sqrt{\xi^{-2}-1}| + \Psi_{0}+\frac{1}{2}\ln |t|,
\\
\gamma&=&\frac{1}{2}\ln \left|\frac{1}{2}
\left(
\frac{\xi^{-1}}{\sqrt{\xi^{-2}-1}}\pm 1\right)\right|
+\lambda,
\end{eqnarray}
where the solution is valid for $ \xi^{2}<1$.
We can assume $x>0$ in this case.

\subsection{Flatness and topology of the solutions}

As indicated in the introduction, our aim in deriving the
solutions of the previous subsection is to study the simplest case of
self-similar cylindrical collapse. However it transpires that these
solutions do not represent collapsing gravitational waves. The
solutions are flat everywhere except along the axis, and thus the
solutions correspond either to part of Minkowski spacetime, or to a line
conical singularity in flat spacetime: the assumption of self-similarity of the gravitational waves rules out any other possibility. This is a nontrivial result regarding self-similar cylindrical collapse. Furthermore, the form of the flat spacetime metric that emerges demonstrates explicitly that there are cylindrical trapping horizons in Minkowski spacetime, and that $C-$energy is (i) nonunique and (ii) nonzero in Minkowski spacetime. This seriously undermines the interpretation of $C-$energy as the gravitational energy of a cylindrical spacetime.

We can explicitly show that all coordinate components of the Riemann
curvature tensor vanish for the obtained solutions.  This means that the solutions are flat.
Here we show that these solutions are indeed part of the Minkowski
spacetimes.

\subsubsection{Untrapped case}
We here assume that $\nabla^{a}r$ is spacelike, i.e.,
$(t,x)$ corresponds to an untrapped cylinder.
The regular axis condition
implies $ \Gamma \to 0 $
as $\xi\to \infty$. From Eq.~(\ref{eq:Gamma_spacelike_r}),
this is possible only for the upper-sign solution
with $ \lambda=0$.
Hence, the solution becomes
\begin{equation}
\gamma=\frac{1}{2}\ln \left|\frac{1}{2}
\left(\frac{\xi}{\sqrt{\xi^{2}-1}}+ 1\right)\right|.
\label{eq:Gamma_spacelike_r_regular}
\end{equation}
The upper-sign solution for the different choice of $\lambda$
gives a conical singularity with the ratio of an infinitesimal
circle's circumference to its diameter
$\pi e^{-\lambda}$ rather than $\pi$.
We first concentrate on the upper-sign solution with $\lambda=0$ below.

\begin{figure}[htbp]
\begin{center}
\includegraphics[width=0.5\textwidth]{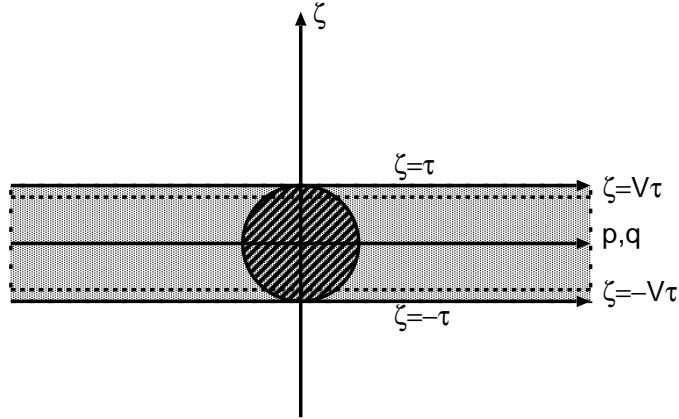}
\caption{\label{fg:tau_const_section}
The constant $\tau$ spacelike hypersurface is
shown in the standard Cartesian coordinates $(\tau,p,q,\zeta)$.
The dark shaded region is untrapped,
while the light shaded region is trapped. The circle shows
the light cone, which is a cylindrical trapping horizon.
The region which is unshaded is not described by the cylindrical
vacuum flat solutions. The dashed lines denote the timelike
planes which are identified with $\zeta=0$.}
\end{center}
\end{figure}

Choosing the coordinates where $\Psi_{0}=0$,
we can write the metric as
\begin{equation}
ds^{2}=\frac{1}{2\sqrt{t^{2}-x^{2}}}(-dt^{2}+dx^{2})+
|t \mp \sqrt{t^{2}-x^{2}}| d\varphi^{2}+|t\pm \sqrt{t^{2}-x^{2}}| dz^{2},
\label{eq:selfsimilar_regular_axis}
\end{equation}
where the solution is valid only for $t^{2}>x^{2}$.
Assuming $t>0$,
through the coordinate transformations
\begin{equation}
T^{2}=t+ \sqrt{t^{2}-x^{2}} \quad \mbox{and}\quad
X^{2}=t- \sqrt{t^{2}-x^{2}},
\label{eq:TX_tx_1}
\end{equation}
or
\begin{equation}
t=\frac{T^{2}+X^{2}}{2} \quad \mbox{and}\quad
x= TX,
\end{equation}
we obtain the following metric for the upper-sign:
\begin{equation}
ds^{2}=-dT^{2}+dX^{2}+X^{2}d\varphi^{2}+T^{2}dz^{2},
\end{equation}
where $0\le X<T$ by construction.
Through another coordinate transformation
\begin{equation}
\tau = T\cosh z, \zeta = T\sinh z,
p=X\cos\varphi ~\mbox{and}~ q=X\sin\varphi
\end{equation}
we finally obtain the usual Minkowski spacetime
in the standard Cartesian coordinates
\begin{equation}
ds^{2}=-d\tau^{2}+dp^{2}+dq^{2}+d\zeta^{2},
\label{eq:Minkowski_metric}
\end{equation}
where $\tau^{2}>p^{2}+q^{2}+\zeta^{2}$ and hence
the solution covers the inside of the light cone
$\tau^{2}=p^{2}+q^{2}+\zeta^{2}$.
This region is shown as the dark shaded disk in
Fig.~\ref{fg:tau_const_section}, where the
constant $\tau$ spacelike hypersurface is plotted.
The $z$-axis, i.e. $x=0$ in $(t,x,\varphi,z)$ coordinates,
is transformed to the $\zeta$-axis, i.e.,
$p=q=0$ in $(\tau,p,q,\zeta)$ coordinates.
The null hypersurface $t^{2}=x^{2}$ in $(t,x,\varphi,z)$ coordinates
is transformed to the light cone $\tau^{2}=p^{2}+q^{2}+\zeta^{2}$
in the standard Cartesian coordinates.

For the lower-sign solution,
the metric is written as
\begin{equation}
ds^{2}=-dT^{2}+dX^{2}+T^{2}d\varphi^{2}+X^{2}dz^{2},
\label{eq:nontrivial_Minkowski_1}
\end{equation}
where $0\le X <T$.
Through another coordinate transformation
\begin{equation}
 \tau=T\cosh\varphi, \zeta=T\sinh \varphi, p =X\cos z
~\mbox{and}~ q =X\sin z,
\end{equation}
we finally obtain the metric
\begin{equation}
ds^{2}=-d \tau^{2}+dp^{2}+dq^{2}+d \zeta ^{2},
\end{equation}
where $\tau^{2}>p^{2}+q^{2}+\zeta^{2}$.
This is also the Minkowski spacetime in the standard Cartesian
coordinates but with nontrivial topology.
In the original $(t,x,\varphi,z)$ coordinates, $\varphi=2\pi$
is identified with $\varphi=0$. This results in the
identification between the two timelike hypersurfaces
$\zeta=0$ and $\zeta=V \tau$, where $V=\tanh 2\pi$.
The latter timelike hypersurface is shown as a dashed line in
Fig.~\ref{fg:tau_const_section}.
More precisely, the point $(\tau, p, q, 0)$ is identified with
the point $(\tau/\sqrt{1-V^{2}}, p,q, \tau V/\sqrt{1-V^{2}})$.
Then, there appears a timelike geodesic with infinite spatial acceleration
in an approach to the spacelike line
$\tau=\zeta=0$, which will be
described in detail in the appendix. The appendix also serves to clarify the nature of the topological identifications made here.
On the other hand, a circle identification has not been imposed on $z$,
implying that the two-dimensional $pq$ plane consists of
covering planes folded infinitely many times.
The $z$-axis is again transformed to the $\zeta$-axis, i.e., $p=q=0$.

We note that here 
the (original) cylinders of symmetry $(t,x)=(t_0,x_0)$ with $t_0,x_0$ both constant have the following representation in Minkowski coordinates $(\tau,p,q,\zeta)$:
\begin{equation} \tau^2-\zeta^2 = t_0+\sqrt{t_0^2-x_0^2},\qquad p^2+q^2 = t_0-\sqrt{t_0^2-x_0^2}.
\end{equation}

\subsubsection{Trapped case}
We then assume that $\nabla^{a}r$ is timelike, i.e.,
$(t,x)$ corresponds to a trapped cylinder.
Also for this case, choosing
$\Psi_{0}=0$ and $\lambda=0 $,
we obtain
\begin{equation}
ds^{2}=\frac{1}{2\sqrt{x^{2}-t^{2}}}(-dt^{2}+dx^{2})+
|x\mp \sqrt{x^{2}-t^{2}}|  d\varphi^{2}
+|x\pm \sqrt{x^{2}-t^{2}}|  dz^{2},
\label{eq:selfsimilar_outside_light_cylinder}
\end{equation}
where the solution is valid only for $x^{2}>t^{2}$.
By implementing coordinate transformations similar to those used in the untrapped case, we obtain for the lower-sign solution
\begin{equation}
ds^{2}=-dT^{2}+dX^{2}+X^{2}d\varphi^{2}+T^{2}dz^{2},
\end{equation}
where $0<T<X$. This is identical with the
upper-sign solution for the untrapped region and
hence transformed to the standard Cartesian coordinates
$(\tau,p,q,\zeta)$, where the solution covers the region
$\zeta^{2}<\tau^{2}<p^{2}+q^{2}+\zeta^{2}$.
This corresponds to the intersection of the outside of light cone
$\tau^{2}=p^{2}+q^{2}+\zeta^{2}$ and the timelike portion sandwiched
by two planes $\tau=\pm \zeta$.
This region is shown as a light shaded region
in Fig.~\ref{fg:tau_const_section}.
The topology of the spacetime is trivial.

For the upper-sign solution, we obtain
\begin{equation}
ds^{2}=-dT^{2}+dX^{2}+T^{2}d\varphi^{2}+X^{2}dz^{2},
\label{eq:nontrivial_Minkowski_2}
\end{equation}
where $0<T<X$. This is identical with the lower-sign solution
for the untrapped region and hence transformed to the
standard Cartesian coordinates
$(\tau,p,q,\zeta)$, where the solution covers the region
$\zeta^{2}<\tau^{2}<p^{2}+q^{2}+\zeta^{2}$, i.e.,
the intersection of the outside of the light cone
and the timelike portion sandwiched
by two planes $\tau=\pm \zeta$.
We should note that the topology is nontrivial because
$\varphi$ is circularly identified while $z$ is not.

We note that in the trapped case the (original) cylinders of symmetry $(t,x)=(t_0,x_0)$ with $t_0,x_0$ both constant have the following representation in Minkowski coordinates $(\tau,p,q,\zeta)$:
\begin{equation} \tau^2-\zeta^2 = x_0-\sqrt{x_0^2-t_0^2},\qquad p^2+q^2 = x_0+\sqrt{x_0^2-t_0^2}.
\end{equation}

\vspace{1cm}

Hence, for $\lambda=0$, the union of
the upper-sign solution for the untrapped region
and the lower-sign solution for the trapped region
describes the timelike portion of the Minkowski
spacetime sandwiched by two light planes
when the two solutions are
matched on the light cone $\tau^{2}=p^{2}+q^{2}+\zeta^{2}$.

The union of the lower-sign solutions for the untrapped region
and the upper-sign solution for the trapped region
also describes the same region of
the Minkowski spacetime when they are
matched on the light cone $\tau^{2}=p^{2}+q^{2}+\zeta^{2}$
but with nontrivial topology.
The solutions with nonvanishing $\lambda$ will have an
additional conical singularity.

This solution without the conical singularity
is quite analogous to the Milne universe solution,
which is also part of the Minkowski spacetime. However, the present
solution is somewhat different from the Milne universe in the following
respect. Recall that observers with constant spatial coordinates
run radially outward with a constant speed and they do so homogeneously
and isotropically in the Minkowski spacetime. On the other hand,
in the present solution, observers with constant spatial coordinates
run with a constant speed but only in the direction parallel to the
axis and do not in the two perpendicular ones. We shall call the present
solution the cylindrical Milne solution in this paper.

\subsection{$C$ energy and trapping horizon in the Minkowski spacetime}

$C-$energy was introduced 
in~\cite{Thorne:1965} as a tool with which cylindrical
spacetimes may be analyzed. It has several interesting and useful features: It is covariant
and is associated with a conserved flux vector;
it has the correct Newtonian limit, the mass per specific length
of the cylinder~\cite{Hayward:2000};
it is propagated by Einstein-Rosen
waves. Thus it is a candidate for
``\textit{the energy of whole-cylinder-symmetric spacetimes}'' (the phrase appears in
quotation marks in~\cite{Thorne:1965}, p.251) and a later study
refers to $C-$energy as
``gravitational energy per specific length''~\cite{Hayward:2000}. As a particular
application, $C-$ energy has been used to 
investigate the fate of an infinitesimally
thin cylindrical shell composed of
counter-rotating dust particles by Apostolatos and
Thorne~\cite{Apos:1992}, and
later by one of the present author (KN) and his
collaborators~\cite{Nakao:2008}:
it should be stressed that the conclusions in these two papers
do not agree with each other although both rely on the properties of
the $C-$energy.
A further criterion that
should be satisfied by a candidate for
gravitational energy of any form is that it
should vanish in the absence of a gravitational field,
i.e.\ in flat spacetime.
It transpires however that the cylindrical
representations of flat spacetime we have
found above show that the $C-$energy \textit{does not}
always vanish in this case.

For the Einstein-Rosen waves written in the forms of
Eqs.~(\ref{eq:metric_t_x_x}) and~(\ref{eq:metric_t_x_t}) ,
the $C$ energy reduces to the following simple forms:
\begin{equation}
E=\frac{1}{8}(1-e^{-2\gamma})
\quad \mbox{and}\quad
E=\frac{1}{8}(1+e^{-2\gamma}),
\end{equation}
respectively.
For simplicity, we discuss the cylindrical Milne solution
with trivial topology, which is given by pasting
the upper-sign solution of
Eq.~(\ref{eq:selfsimilar_regular_axis}) and
the lower-sign solution of
Eq.~(\ref{eq:selfsimilar_outside_light_cylinder})
on the null hypersurface $t^{2}=x^{2}$.
For these metrics, we obtain respectively
\begin{equation}
E=\frac{1}{8}\frac{t-\sqrt{t^{2}-x^{2}}}
{t+\sqrt{t^{2}-x^{2}}},\quad\mbox{and}\quad
E=\frac{1}{8}\frac{x+\sqrt{x^{2}-t^{2}}}
{x-\sqrt{x^{2}-t^{2}}}.
\end{equation}
This is rewritten in both cases as
\begin{equation}
E=\frac{1}{8}\frac{p^{2}+q^{2}}{\tau^{2}-\zeta^{2}},
\end{equation}
in terms of the standard Cartesian coordinates.
Since these metrics are those for part of
the Minkowski spacetime, it
can have nonvanishing $C$ energy.

However, if we write the metric of the Minkowski spacetime in the
standard cylindrical coordinates,
we have $\gamma=\psi=0$ in Eq.~(\ref{eq:metric_t_x_x}) and hence $E=0$.
This result questions the physical interpretation of the $C$ energy. Indeed,
the trick is in the choice of the two commuting Killing vectors,
or equivalently, the choice of the cylinders.
If we take $(\xi_{(1)},\xi_{(2)})= (\partial/\partial
\varphi,\partial/\partial z)$, then $\rho$, $\ell$ and $r$ are
calculated as
\begin{eqnarray}
\rho^{2}&=& |t-\sqrt{t^{2}-x^{2}}|=p^{2}+q^{2}, \\
\ell^{2}& =& |t+\sqrt{t^{2}-x^{2}}|=\tau^{2}-\zeta^{2}, \\
r^{2}&=&x^{2}=(\tau^{2}-\zeta^{2})(p^{2}+q^{2})
\label{eq:complicated_cylinder},
\end{eqnarray}
where $t^{2}>x^{2}$ or $\tau^{2}>p^{2}+q^{2}+\zeta^{2}$.
Hence, the regular axis condition is satisfied
and we obtain nontrivial $C$ energy. Instead, if we take
$(\xi_{(1)},\xi_{(2)})= (\partial/\partial \varphi,\partial/\partial
\zeta)$, then we have $\rho^{2}=p^{2}+q^{2}$,
$\ell=1$ and $r^{2}=p^{2}+q^{2}$ and obtain vanishing $C$ energy.
We should also note that the $z$-axis in the former is transformed
to $\zeta$-axis in the latter. This means that the definition of
$C$ energy is ambiguous for the same axis in the same cylindrically
symmetric spacetime unless a pair of two commuting Killing vectors
are fully specified. Thorne~\cite{Thorne:1965} has noted the lack of
uniqueness in the definition of $C$ energy in the case of
unpolarized cylindrical spacetimes, for which the Killing vectors
$(\xi_{(1)}, \xi_{(2)})$ are not orthogonal. Nonuniqueness in the
unpolarized case is related to the loss of invariance of the
spacetime under reflections through any plane either containing the
axis or perpendicular to it. However as we see in the present case,
nonuniqueness can remain even in the polarized case when there is
more than one choice of the azimuthal and translational Killing fields.

This clearly gives rise to a question about the interpretation of
$C$ energy as ``{\em gravitational energy per
specific length}''~\cite{Hayward:2000}, given that it may be nonzero in the absence of
a gravitational field. But as we have pointed out,
$C$ energy has many attractive and useful features, and so perhaps
the most natural question to ask at this point is if there exists an
alternative definition that would have the additional feature of
vanishing for {\em any} cylindrical slicing of Minkowski spacetime.
We hope to address this question in future work.

\begin{figure}[htbp]
\begin{center}
\includegraphics[width=0.5\textwidth]{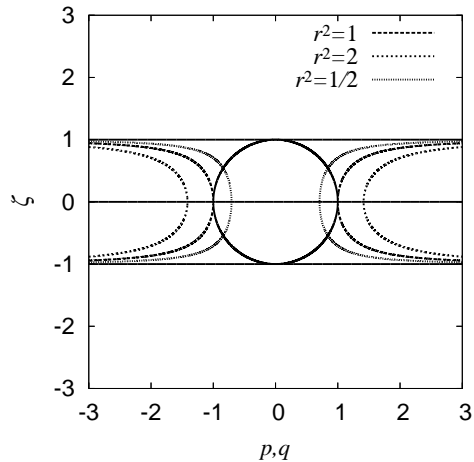}
\caption{\label{fg:cylindrical_surfaces}
The constant $\tau$ spacelike hypersurface is
shown in the standard Cartesian coordinates
$(\tau,p,q,\zeta)$, where $\tau$ is chosen to be unity.
The short-dashed, dashed and long dashed lines denote the
intersection of the
constant $r$ hypersurfaces given by
 Eq.~(\ref{eq:complicated_cylinder})
with the constant $\tau$ hypersurface.
}
\end{center}
\end{figure}

Although the uniqueness of the $C$ energy may be recovered by {\em
specifying} the pair of Killing vectors, it is still true that
the null hypersurface $t^{2}=x^{2}$ in the
original coordinates or the light cone
$\tau^{2}=p^{2}+q^{2}+\zeta^{2}$ in the standard Cartesian coordinates
gives a cylindrical trapping horizon. The inside of the light cone
$\tau^{2}>p^{2}+q^{2}+\zeta^{2}$ is untrapped, while
the outside of the light cone,
i.e., $\zeta^{2}<\tau^{2}<p^{2}+q^{2}+\zeta^{2}$ is trapped.
The constant $r$ hypersurfaces, given by
Eq.~(\ref{eq:complicated_cylinder}),
are shown in Fig.~\ref{fg:cylindrical_surfaces}
on the constant $\tau$ hypersurface.
On the other hand,  with the pair
$(\partial/\partial \varphi, \partial /\partial \zeta)$
of Killing vectors,
the constant $r$ hypersurfaces given by
\begin{equation}
p^{2}+q^{2}=r^{2}
\end{equation}
are all timelike.
See also~\cite{Senovilla:1997} for trapped surfaces in the
Minkowski spacetime.

In cylindrically symmetric spacetimes, a
trapping horizon is defined as a hypersurface foliated by marginally
trapped cylinders and hence will not be closed in general.
Since they are not closed, it will not imply the existence of spacetime
singularity. Thus,
this example is a lesson that we cannot reasonably identify a
trapping horizon with a black hole horizon for cylindrically
symmetric spacetimes~\cite{Hayward:2000,Wang:2003}.
On the other hand, we could have closed trapped surfaces if we change the
identification. In that case, we may encounter a sort of singularity
as described in the appendix.

\section{More general self-similar Einstein-Rosen waves}

In the previous two sections, we showed that the only
self-similar cylindrically symmetric vacuum spacetimes comprise flat
spacetimes, possibly with line conical singularities along the
axis. These are trivial examples of self-similar spacetimes, and so
while these may act as (intermediate) asymptotic endstates of certain
more general cylindrical configurations - and indeed a line conical singularity in an otherwise flat spacetime can result from the complete collapse of cylindrical null dust - it is clear that this does not lend any weight to the self-similarity hypothesis in cylindrical symmetry. It transpires however that part of the reason for this is that we have looked only at a quite restrictive class of self-similar cylindrical spacetimes. Dropping the assumption that the homothetic vector field is orthogonal to the cylinders of symmetry yields some interesting results which we describe here.

\subsection{Two-parameter family of solutions}

In this section, we concentrate on the spacelike $\nabla^{a}r$ case,
where the region is untrapped.
In this case, Eqs.~(\ref{eq:wave_spacelike}), (\ref{eq:gammax_spacelike}) and
(\ref{eq:gammat_spacelike}) give a complete set of governing equations.
Among them, Eq.~(\ref{eq:wave_spacelike}) gives the dynamics of $\psi$ and
the other equations determine the derivatives of $\gamma$.
For self-similar solutions with a cylindrical homothetic vector, $\psi$ is given
by Eq.~(\ref{eq:selfsimilar_psi_spacelike}).
Note that each of two terms on the right-hand side of
Eq.~(\ref{eq:selfsimilar_psi_spacelike}) is a solution of
 Eq.~(\ref{eq:wave_spacelike}) - provided that the first term $\Psi$ is taken to be a solution of Eq.~(\ref{eq:selfsimilar_wave}).
Since Eq.~(\ref{eq:wave_spacelike}) is linear, this means that the
arbitrary linear combination of the two terms gives a solution of
Eq.~(\ref{eq:wave_spacelike}). Moreover, we can assume the similar
form also for $\gamma$ from
Eq.~(\ref{eq:selfsimilar_gamma_spacelike}). From this consideration,
we assume the following form for $\psi$ and $\gamma$:
\begin{eqnarray}
\psi&=& a \tilde{\Psi}(\xi)+b \frac{1}{2} \ln |x|,
\label{eq:extended_selfsimilar_ansatz_psi}\\
\gamma &=& c \tilde{\Gamma}(\xi) + d \frac{1}{2}\ln |x|.
\end{eqnarray}
 From Eqs.~(\ref{eq:wave_spacelike}), ~(\ref{eq:gammax_spacelike})
and ~(\ref{eq:gammat_spacelike}), we get the following
ordinary differential equations
for $\tilde{\Psi}$ and $\tilde{\Gamma}$:
\begin{eqnarray}
&& (\xi^{2}-1)\tilde{\Psi}''+\xi \tilde{\Psi}'=0,
\label{eq:extended_selfsimilar_wave}\\
&& -c \xi \tilde{\Gamma}'+\frac{d}{2}=
\left(a \xi \tilde{\Psi}'-\frac{b}{2}\right)^{2}+ a^{2} \tilde{\Psi}'^{2},
\label{eq:extended_Gamma1}\\
&& c\tilde{\Gamma}'=-2 \left(a \xi \tilde{\Psi}'-\frac{b}{2}\right)a\tilde{\Psi}'.
\label{eq:extended_Gamma2}
\end{eqnarray}
 From Eq.~(\ref{eq:extended_selfsimilar_wave}), we get
\begin{equation}
\tilde{\Psi}'= \frac{\tilde{\Psi}_{1}}{\sqrt{|\xi^{2}-1|}}.
\label{Psi1}
\end{equation}
We can always assume $\tilde{\Psi}_{1}=1/2$ because of the factor $a$
in Eq.~(\ref{eq:extended_selfsimilar_ansatz_psi}).
Eliminating $\tilde{\Gamma}'$ from Eqs.~(\ref{eq:extended_Gamma1})
and (\ref{eq:extended_Gamma2}), using Eq.~(\ref{Psi1})
and putting $\tilde{\Psi}_{1}=1/2$,
we obtain
\begin{equation}
\mbox{sign}(1-\xi^{2}) a^{2}+b^{2}=2d.
\end{equation}
Then, for $\xi^{2}>1$, the solution is given by
\begin{eqnarray}
\psi&=& a \frac{1}{2}\ln |\xi+\sqrt{\xi^{2}-1}|+b \frac{1}{2}\ln x
+\tilde{\Psi}_{0}, \\
\gamma&=&
a \frac{1}{2}\left[-a \ln \sqrt{\xi^{2}-1}+ b \ln |\xi+\sqrt{\xi^{2}-1}|
\right]
+d \frac{1}{2}\ln |x|+\tilde{\Gamma}_{0},
\end{eqnarray}
where we omit the lower-sign solution because the sign can be absorbed
into the sign of $a$.

To be more specific, we assume
that the axis is regular or conically singular at least,
which implies that $\gamma$ approaches a finite value
for $t>0$ and $x\to 0$.
This condition strongly restricts the parameters.
This limit corresponds to $\xi\to \infty$ and $x\to 0$, where
$\xi^{-1}$ and $x$ approach zero independently.
Hence, the condition on the axis implies
$a=b$, $d=0$ and a finite value for $\tilde{\Gamma}_{0}$.
Putting $a=b=2\kappa$ and $\tilde{\Gamma}_{0}=\lambda$, we obtain the following
solution with a regular or conically singular axis:
\begin{eqnarray}
\psi&=&  \kappa \left[ \ln (\xi+\sqrt{\xi^{2}-1})+\ln |x|
\right],
\label{eq:The_solution_psi}\\
\gamma&=& 2\kappa^{2}\ln \left|\frac{1}{2}
\left(\frac{\xi}{\sqrt{\xi^{2}-1}}+ 1\right)\right|+\lambda,
\label{eq:The_solution_gamma}
\end{eqnarray}
where $\tilde{\Psi}_{0}$ is eliminated in use of the scaling
freedom of $t$ and $x$.
These solutions are parametrized by $\kappa$
and $\lambda$.
Note that the solution is self-similar in the sense discussed so far
if $\kappa=1/2$ and
the axis is regular if and only if $\lambda=0$.
As we will see later, the spacetime is
nonflat  except for $\kappa=0$ and $1/2$.

For later convenience, we write down the line element
explicitly both in $(t,x,\varphi,z)$ and $(T,X,\varphi,z)$ coordinates as,
\begin{eqnarray}
ds^2&=&\frac{(t+\sqrt{t^2-x^2})^{2\kappa(2\kappa-1)}}{2^{4\kappa^2}(t^2-x^2)^{2\kappa^2}}
e^{2\lambda}(-dt^2+dx^2)
+\frac{x^2}{(t+\sqrt{t^2-x^2})^{2\kappa}}d\varphi^2
+(t+\sqrt{t^2-x^2})^{2\kappa}dz^2 \nonumber \\
&=&\frac{T^{4\kappa(2\kappa-1)}}{(T^2-X^2)^{4\kappa^2-1}}e^{2\lambda}(-dT^2+dX^2)
+\frac{X^2}{T^{2(2\kappa-1)}}d\varphi^2
+T^{4\kappa}dz^2,
\label{eq:metric}
\end{eqnarray}
where $T$ and $X$ are given by Eq.~(\ref{eq:TX_tx_1}).
The original domain of the solution is given by $0\le x<t<\infty$ and
this is mapped to $0\le X<T<\infty$.
It is clear that for $\kappa=0$ the solution reduces to the
Minkowski spacetime with a regular ($\lambda=0$) or a conically
singular ($\lambda\ne 0$) axis.
Moreover, we can easily
find that for $\kappa=-1/2$ this reduces to a Kasner
solution with a regular ($\lambda=0$) or a conically singular
($\lambda\ne 0$) axis.

\subsection{Non cylindrical homothetic vector}
To make it clear whether the solutions obtained above have
some kind of self-similarity,
we shall consider the scaling transformation $\bar{t}=At$, $\bar{x}=Ax$,
$\bar{\varphi}=\varphi$ and $\bar{z}=z$.
Through this transformation, $\psi$ and $\gamma$ transform as follows
\begin{equation}
\bar{\psi} =  \psi+ 2\kappa \ln A \quad \mbox{and} \quad
\bar{\gamma} = \gamma .
\end{equation}
Then, the metric components $g_{tt}$, $g_{xx}$, $g_{\varphi\varphi}$
and $g_{zz}$ transform as follows:
\begin{equation}
\bar{g}_{tt}= A^{2(1-\kappa)}g_{tt},\quad
\bar{g}_{xx}= A^{2(1-\kappa)}g_{xx},\quad
\bar{g}_{\varphi\varphi}= A^{2(1-\kappa)}g_{\varphi\varphi}\quad
\mbox{and}\quad
\bar{g}_{zz}= A^{2\kappa} g_{zz}.
\end{equation}
Therefore, for $\kappa\ne 1$, if we define a vector field ${\bf v}$ as
\begin{equation}
{\bf v}:=\frac{1}{1-\kappa}t\frac{\partial }{\partial t}+
\frac{1}{1-\kappa}x\frac{\partial }{\partial x},
\end{equation}
we can write
\begin{equation}
{\cal L}_{{\bf v}} g_{tt}= 2g_{tt}, \quad
{\cal L}_{{\bf v}} g_{xx}= 2g_{xx}, \quad
{\cal L}_{{\bf v}} g_{\varphi\varphi}= 2g_{\varphi\varphi}\quad
\mbox{and} \quad
{\cal L}_{{\bf v}} g_{zz}= \frac{2 \kappa}{1-\kappa}g_{zz}.
\end{equation}
It is clear that
if and only if $\kappa=1/2$, the vector field ${\bf v}$ is a homothetic
vector.

However, if we instead scale the coordinates as $\bar{t}=At$, $\bar{x}=Ax$,
$\varphi=\varphi$ and $\bar{z}=A^{1-2\kappa}z$,
the metric components transform as follows:
\begin{equation}
\bar{g}_{\mu\nu}= A^{2(1-\kappa)}g_{\mu\nu}.
\end{equation}
For $\kappa\ne 1$, this implies that for the vector field ${\bf w}$ given by
\begin{equation}
{\bf w}:=\frac{1}{1-\kappa}t\frac{\partial }{\partial t}+
\frac{1}{1-\kappa}x\frac{\partial }{\partial x}
+\frac{1-2\kappa}{1-\kappa}z\frac{\partial}{\partial z},
\end{equation}
we obtain
\begin{equation}
{\cal L}_{{\bf w}}g_{\mu\nu}=2 g_{\mu\nu}.
\end{equation}
Therefore, ${\bf w}$ is a homothetic vector and hence
the spacetime described by this solution is self-similar.
The homothetic vector ${\bf w}$ is not cylindrical
if $\kappa\ne 1/2$.

For $\kappa=1$,
if we instead define ${\bf W}$ as
\begin{equation}
{\bf W}:= t\frac{\partial }{\partial t}+
x\frac{\partial }{\partial x}
-z\frac{\partial}{\partial z},
\end{equation}
we obtain
\begin{equation}
{\cal L}_{{\bf W}}g_{ab}=0,
\end{equation}
and therefore ${\bf W}$ is a Killing vector. This is independent from
the azimuthal and the translational Killing vectors.

\section{Physical interpretation of self-similar Einstein-Rosen waves}

In the previous section we have derived a two-parameter family of
self-similar solutions, where the homothetic vector is not
cylindrical in general.
As we have seen, the two-parameter family of solutions
includes a Minkowski ($\kappa=0$),
cylindrical Milne ($\kappa=1/2$) and Kasner ($\kappa=-1/2$) solutions.
In this section we present the
physical interpretation of the family of solutions
for general values of $\kappa$, based on the
analysis of spacetime singularities, infinities
and analytical extensions. We show that these solutions describe
physically interesting nonlinear phenomena of gravitational waves.

\subsection{Curvature invariant, $C-$energy and areal radius}

For the two-parameter family of solutions obtained here,
the Kretschmann invariant $I:=
R^{\mu\nu\rho\sigma}R_{\mu\nu\rho\sigma}$ is calculated to give
\begin{eqnarray}
I=&&2^{4+8 \kappa^2}\kappa^2(1-2\kappa)^2
e^{-4\lambda}(t +\sqrt{t ^2-x^2})^{-2(4\kappa^2-2\kappa+1)}
(t ^2-x^2)^{4\kappa^2-3/2} \nonumber \\
&\times&\left[(1+2\kappa)(1-\kappa)t
+ (2-\kappa+2 \kappa^2 )\sqrt{t ^2-x^2}\right] \nonumber \\
&=& 2^{6}\kappa^{2}(1-2\kappa)^{2}e^{-4\lambda}T^{-4(4\kappa^{2}-2\kappa+1)}
(T^{2}-X^{2})^{8\kappa^{2}-3}
[3T^{2}-(4\kappa^{2}-2\kappa+1)X^{2}].
\end{eqnarray}
This is identically zero and hence
the spacetime is flat if and only if $\kappa=0$ or $1/2$.
For $T^{2}=X^{2}$ the invariant is diverging
if $0<\kappa^{2}<3/8$ and $\kappa^{2}\ne 1/4$,
while it is finite if $\kappa^{2}\ge 3/8$ or $\kappa=0,\pm 1/2 $.
The invariant is finite at the $z$-axis or $x=0$
if $0<t<\infty$. The invariant at the axis is vanishing
even for $0\le t<\infty$
if $\kappa=0$, $\kappa=1/2$ or $\kappa>1$, while it is diverging
at $t=0$ if $\kappa<0$, $0<\kappa<1/2$ and $1/2<\kappa<1$.
Only for $\kappa=1$, it is nonzero and finite for $0\le t <\infty$
on the axis.
Since the solution is vacuum, the Riemann tensor reduces
to the Weyl tensor. Hence, in a intuitive sense, we can
regard the Kretschmann invariant as the field strength
of the pure gravitational field and/or gravitational
waves, although this may be negative for $\kappa<-1/2$.

The $C$ energy for this solution is calculated to give
\begin{equation}
E=\frac{1}{8}\left[1-e^{-2\lambda}\left\{\frac{1}{2}
\left(\frac{t}{\sqrt{t^{2}-x^{2}}}+1\right)\right\}
^{-4\kappa^{2}}\right]=
\frac{1}{8}\left[1-e^{-2\lambda}\left(
\frac{T^{2}-X^{2}}{T^{2}}\right)
^{4\kappa^{2}}\right].
\end{equation}
$E=(1-e^{-2\lambda})/8$ at the axis $x=0$. This is nonzero
if and only if $\lambda\ne 0$, i.e., at the conical singularity.
Whether $\lambda$ is zero or not,
$E=1/8$ for $t^{2}=x^{2}$
if only $\kappa\ne 0$, suggesting a cylindrical trapping horizon.
For $\kappa=0$, the $C$ energy is identically
given by $E=(1-e^{-2\lambda})/8$ and hence constant
in the whole region described by this solution.
As already mentioned, $E$ is vanishing for $\kappa=\lambda=0$,
but is not for $\kappa=1/2$ and $\lambda=0$, although
both correspond to a flat geometry.
The $C$ energy is unchanged by flipping the sign of $\kappa$.

The areal radius $r$ of the cylinder can be calculated as
\begin{equation}
r^{2}=x^{2}=T^{2}X^{2}.
\end{equation}
Hence, the $r=$const surface is given by a hyperboloid in the $TX$ plane.

\subsection{Analytical extension and global structure}

We recall that the original domain of the solutions
is mapped to $0\le X<T<\infty $ in $TX$ plane.
The ``event'' $t=x=0$, or equivalently, $T=X=0$
seems to be a $s.p.$ (scalar polynomial) curvature
singularity because of the divergence of the curvature
invariant if $\kappa<1$ and $\kappa\ne 0, 1/2$.
But this is a subtle issue. The proper time
$\sigma $ along a curve on which $X$ vanishes and $z$ is constant
is given by
\begin{equation}
\sigma =e^{\lambda}\int T^{-2\kappa+1}dT.
\end{equation}
We can easily see that $\sigma$ is finite in an approach to
$T=0$ if and only if $\kappa<1$.
If and only if $\kappa\geq 1$, then $T=X=0$ is at a
timelike infinity and hence not within the physical spacetime.
On the other hand, $(T,X)=(\infty,0)$
is at a timelike infinity for $\kappa\le 1$,
while it is a spacetime singularity for $\kappa>1$.

For $t^{2}=x^{2}$, or equivalently, $T^{2}=X^{2}$,
the curvature invariant is diverging as long as
$0<\kappa^2<3/8$ and $\kappa^{2}\ne 1/4$.
We introduce the following null coordinates:
\begin{equation}
U=T-X~~~~~{\rm and}~~~~~V=T+X.
\end{equation}
Then, we have
\begin{equation}
ds^2=-e^{2\lambda}\left(\frac{V+U}{2}\right)^{4\kappa(2\kappa-1)}
\frac{dUdV}{(UV)^{4\kappa^2-1}}
+\left(\frac{V+U}{2}\right)^{2(1-2\kappa)}
\left(\frac{V-U}{2}\right)^{2}d\varphi^2
+\left(\frac{V+U}{2}\right)^{4\kappa}dz^2.
\label{eq:1st-metric}
\end{equation}
The original domain of the solutions is mapped to $0<U\le V<\infty$.
Here we consider radial null geodesics along
which $U$, $\varphi$ and $z$ are constant.
The geodesic equation for these null geodesics is given by
\begin{equation}
\frac{d}{d\omega}\left[\frac{(V+U)^{4\kappa(2\kappa-1)}}{(VU)^{4\kappa^2-1}}
\frac{dV}{d\omega}\right]=0,
\label{eq:affine}
\end{equation}
where $U=$const and $\omega$ is an affine parameter.
Thus, we can find that $V=\infty$ is a null infinity for any $\kappa$.
We can similarly find that $U=0$ can be reached in a finite affine length
along null geodesics with $V$=const
for $0<\kappa^{2}<1/2$, while it is a null infinity for $\kappa^{2}\ge 1/2$.
Below we discuss the cases $\kappa^2=1/2$, $0<\kappa^{2}<1/2$
and $\kappa^2>1/2$, separately.

\subsubsection{$\kappa^{2}=1/2$}
First, we consider the case of $\kappa^2=1/2$, and introduce
\begin{equation}
U=e^{u}~~~~~~{\rm and}~~~~~~V=e^{v}.
\end{equation}
Then, the metric (\ref{eq:1st-metric}) becomes
\begin{equation}
ds^2=-e^{2\lambda}\left(\frac{e^v+e^u}{2}\right)^{2(2\mp \sqrt{2})}dudv+
\left(\frac{e^{v}+e^{u}}{2}\right)^{2(1\mp \sqrt{2})}
\left(\frac{e^{v}-e^{u}}{2}\right)^{2}d\varphi^2
+\left(\frac{e^{v}+e^{u}}{2}\right)^{\pm 2\sqrt{2}}dz^2
\label{eq:2nd-metric}
\end{equation}
for $\kappa=\pm 1/\sqrt{2}$.
The original domain is mapped to $-\infty<u\le v<\infty$ and this
describes a maximal extension because $v=\infty$ and $u=-\infty$
both correspond to null infinities.
In the above coordinate system, there seems to be no singular point.
But as already shown,
the ``event'' $u=v= -\infty$ or $T=X=0$
is a spacetime singularity.
The conformal diagram of the solution is given in
Fig.~\ref{fg:penrose_kappa2gtr1o2_kappalt1}.
There is no cylindrical trapping horizon and
the whole spacetime is untrapped.

\subsubsection{$0<\kappa^{2}<1/2$}
Next, we consider the case of $0<\kappa^2<1/2$. In this case, we introduce
$u$ and $v$ as
\begin{equation}
U=u^{n}~~~~~~{\rm and}~~~~~~V=v^{n},
\label{eq:uv-definition}
\end{equation}
where $n:= 1/[2(1-2\kappa^{2})]>0$.
Then, we have
\begin{equation}
ds^2=-[2(1-2\kappa^{2})^{2}]^{-2}
e^{2\lambda}\left(\frac{V+U}{2}\right)^{4\kappa(2\kappa-1)}
dudv
+\left(\frac{V+U}{2}\right)^{2(1-2\kappa)}
\left(\frac{V-U}{2}\right)^{2}
d\varphi^2
+\left(\frac{V+U}{2}\right)^{4\kappa}dz^2.
\label{eq:3rd-metric}
\end{equation}
The original domain is mapped to $0<u\le v<\infty$.
As we have shown, $v=\infty$ is a null infinity, while $u=0$
is finite. For $0<\kappa^{2}<3/8$ and $\kappa^{2}\ne 1/4$,
the Kretschmann invariant
diverges at $u=0$ and hence $u=0$ corresponds to a null
singularity.
For $3/8\le \kappa^{2}<1/2$, $u=0$ is an at least $C^{2}$ 
extendible null hypersurface,
which coincides with a cylindrical trapping horizon,
and we can discuss the extension beyond this hypersurface.
To examine the affine length of the radial null geodesic with $u=0$,
we should consider
\begin{equation}
\frac{d}{d\omega}\left[(V+U)^{4\kappa(2\kappa-1)}
\frac{dv}{d\omega}\right]=0,
\end{equation}
instead of Eq.~(\ref{eq:affine}).
We can then find that the affine length is infinite
to $v=\infty$ and finite to $v=0$ even along
the null geodesic with $u=0$ for $3/8\le \kappa^{2}<1/2$.

If and only if $n$ is a natural number, the extension
beyond this surface can be analytic and we can naturally
extend the spacetime by Eq.~(\ref{eq:uv-definition}).
The following discussion depends on whether $n$
is odd or even.

If $n=2l+1$ ($l=1,2,\cdots)$, the maximally extended domain is given by
$\{0<v<\infty~\mbox{and}~ -v<u\le v\}$.
On the surface $v=-u$, we have $T=0$ and
hence the Kretschmann invariant diverges.
So this surface corresponds to a spacelike singularity.
There is a cylindrical trapping horizon on $u=0$.
The region $\{0<u\le v<\infty \}$ is untrapped, while
the region $\{0<-u<v<\infty \}$ is trapped.

If $n=2l$ ($l=1,2,\cdots$), the maximally extended spacetime is 
apparently given by
$-\infty<u\le v<\infty$. $u=-\infty$ is a null infinity.
$v=-u=\infty$ is a spacelike infinity.
$u=v=0$ is a spacetime singularity.
$v=u=\pm\infty$ are both timelike infinities.
It is interesting to see the surface $v=-u$.
Noting $T=(v^{n}+u^{n})/2$ and $X=(v^{n}-u^{n})/2$, we find that
the Kretschmann invariant is finite there except for $u=v=0$,
while the areal radius $r$ vanishes.
It turns out that we need to pay close attention to this surface.
To get an insight into this surface, we introduce $t$ and $x$
coordinates, where $u=t-x$ and $v=t+x$, so that we should focus
on the surface $t=0$.
Near this surface, the metric can be written as
\[
 ds^{2}\simeq
[2(1-2\kappa^{2})]^{-2} x^{2\kappa(2\kappa-1)/(1-2\kappa^{2})}
\left[e^{2\lambda}(-dt^{2}+dx^{2})+t^{2}d\varphi^{2}\right]
+x^{2\kappa/(1-2\kappa^{2})}dz^{2}.
\]
It follows from
the identification between $\varphi=0$ and $\varphi=2\pi$
that there is a timelike geodesic
on the $t\varphi$ plane with infinite spatial acceleration,
as shown in the appendix, and hence the spacetime is geodesically incomplete in a sense that 
there is a geodesic which cannot be uniquely extended. Thus, there is no analytical extension beyond this
surface. Then, the structure of the resultant spacetime is similar to
the $n=2l+1$ case, except for that ``singularity'' on the surface $t=0$ is only 
``quasiregular'' in the sense of Ellis and 
Schmidt~\cite{Ellis:1977}.

It should be noted that it is impossible to analytically
extend the spacetime
even beyond $u=0$ even for $3/8<\kappa^2<1/2$ if $n= 1/[2(1-2\kappa^{2})]$
is not an integer. For this case,
the functions $U$ and $V$ are at least twice differentiable but
not $C^\infty$ with respect to $u$ at $u=0$.
The spacetime admits at most $C^{[n]}$ extension beyond the null
surface $u=0$, where $[n]$ is the largest integer which is no
greater than $n$.

The conformal diagrams of the solutions for different values of $\kappa$
are given in Figs.~\ref{fg:penrose_kappa2lt3o8} and
\ref{fg:penrose_integern}.
As we can see in these figures, the case of integer $n$ 
is particularly intriguing in the context of 
gravitational collapse
because these solutions are self-similar,
describe the collapse of gravitational waves
and admit nonsingular
initial data on a spacelike Cauchy surface 
containing both trapped and untrapped cylinders.

\subsubsection{$\kappa^{2}>1/2$}
\begin{figure}[htbp]
\begin{center}
\begin{tabular}{cccc}
\subfigure[]
{\includegraphics[width=0.25\textwidth]
{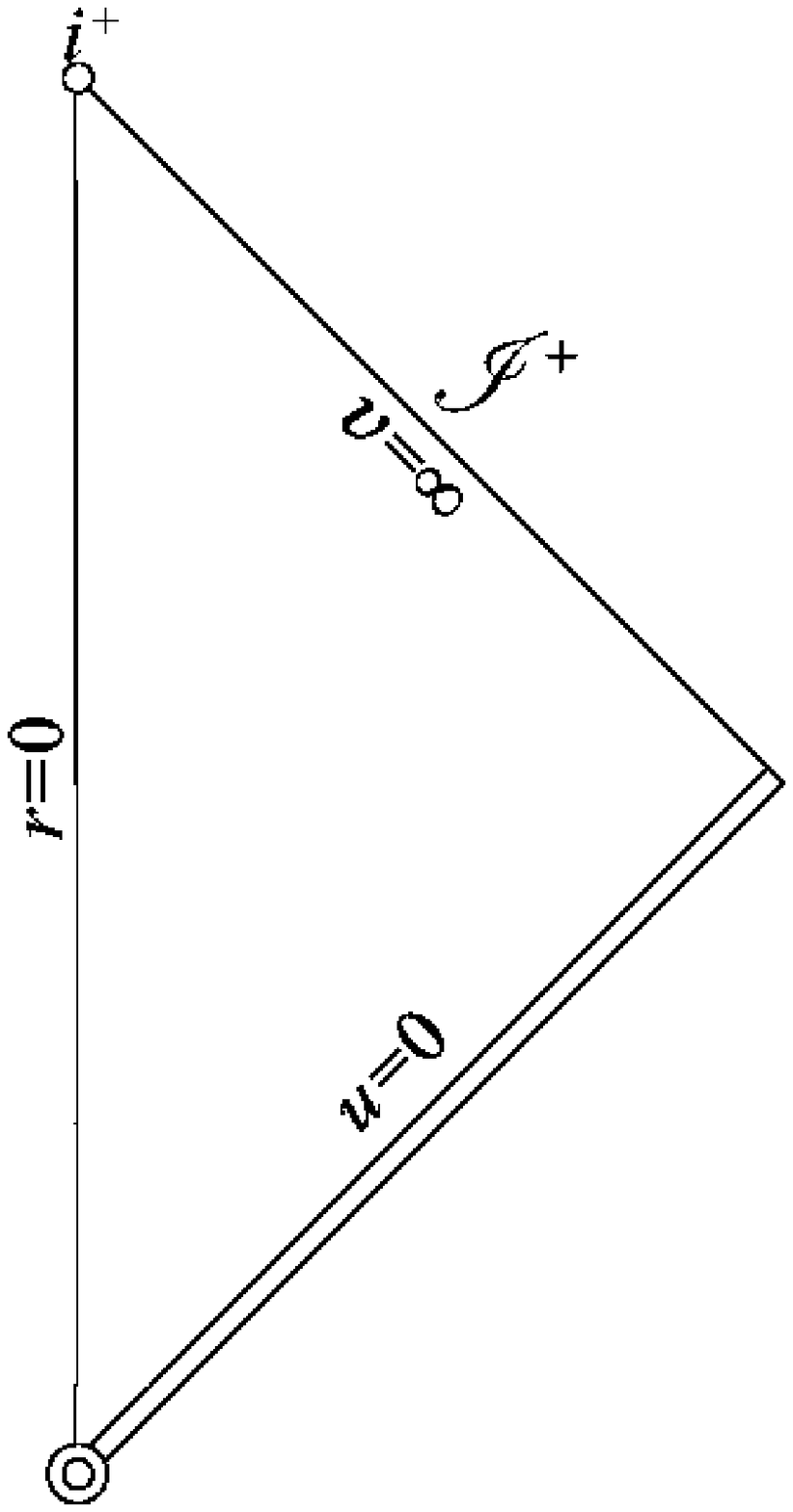}
\label{fg:penrose_kappa2lt3o8}}&
\subfigure[]
{\includegraphics[width=0.4\textwidth]
{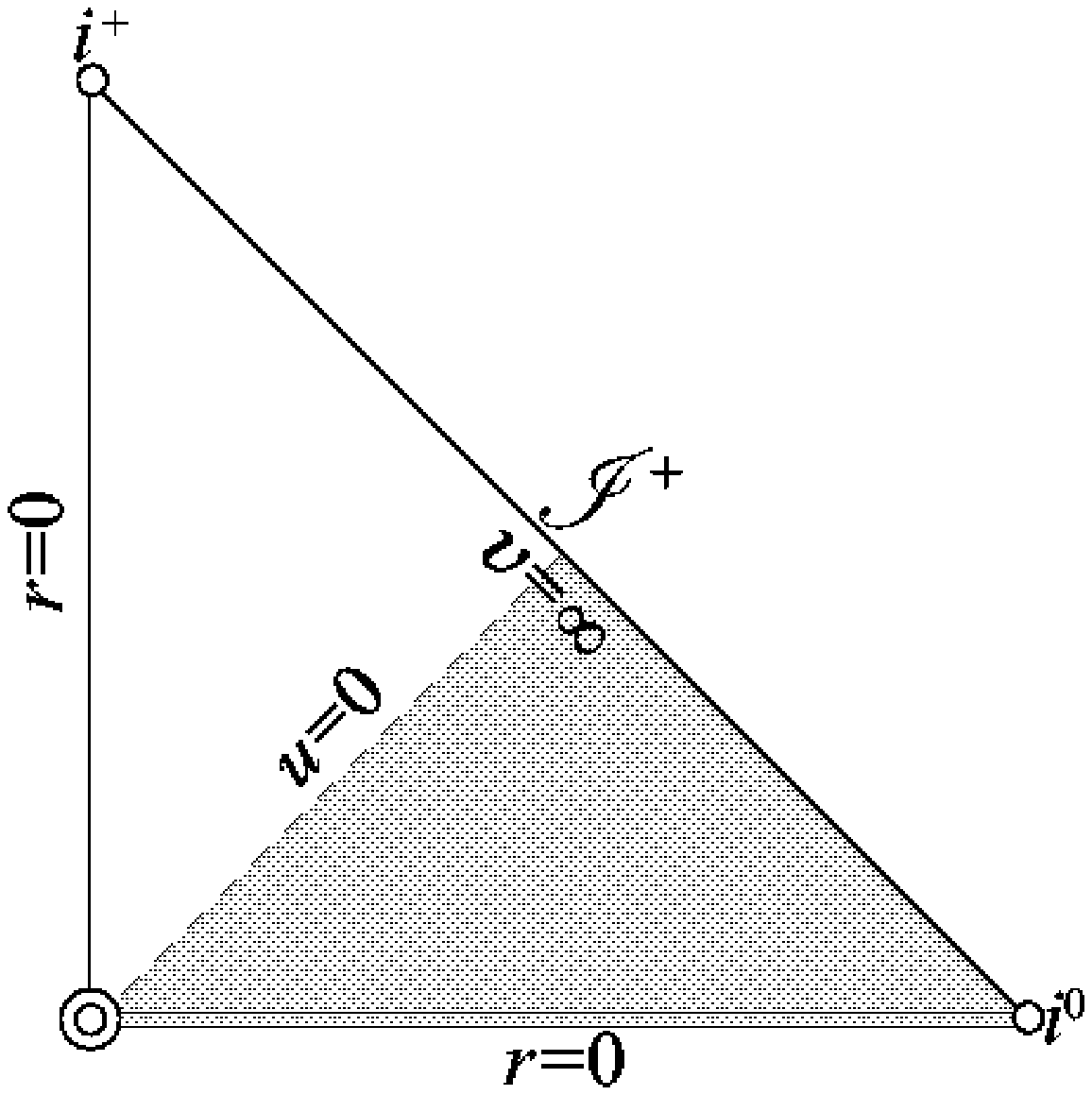}
\label{fg:penrose_integern}} &
\subfigure[]
{\includegraphics[width=0.25\textwidth]
{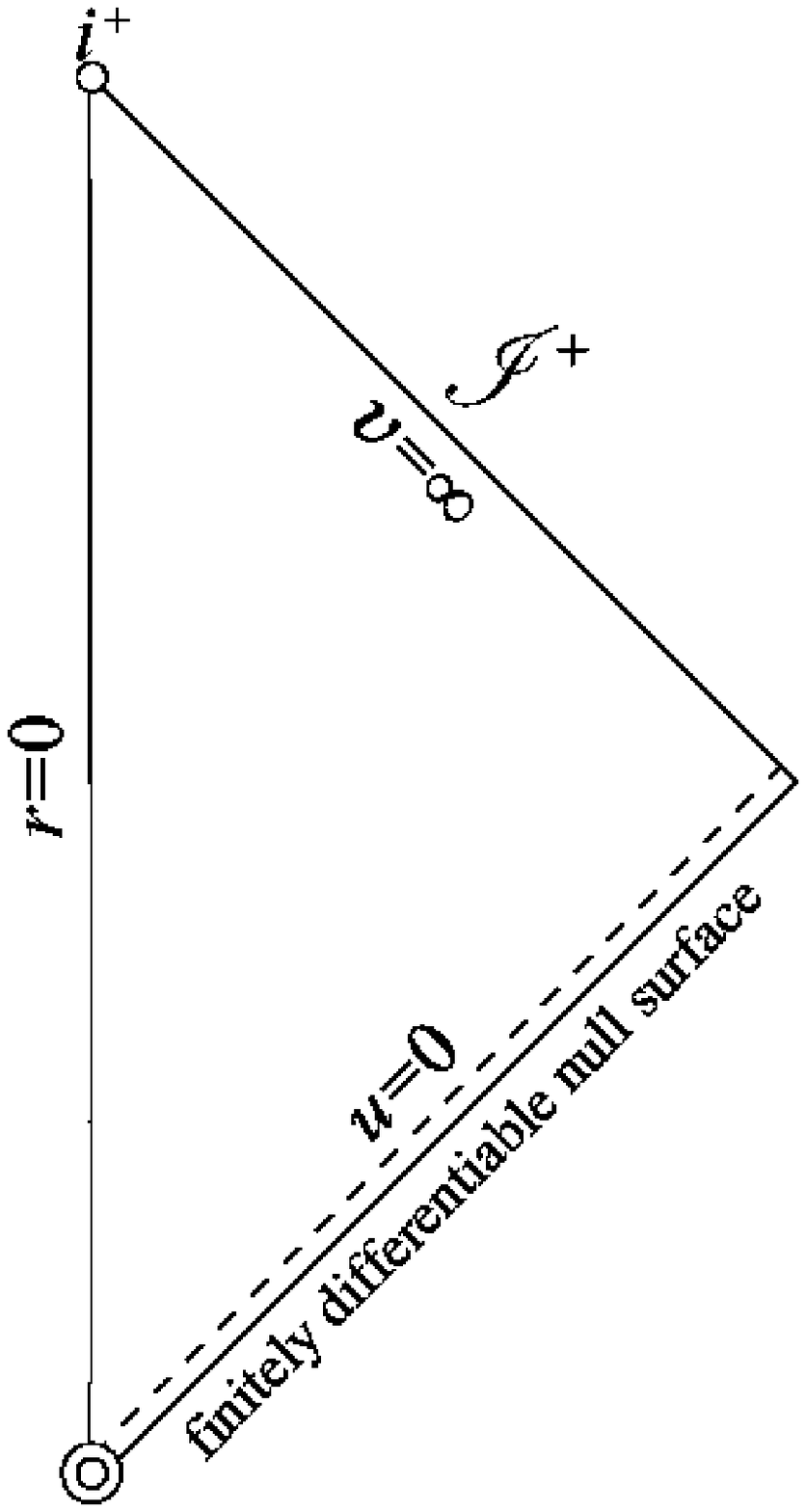}
\label{fg:penrose_positivenonintegern}} \\
\subfigure[]{\includegraphics[width=0.25\textwidth]
{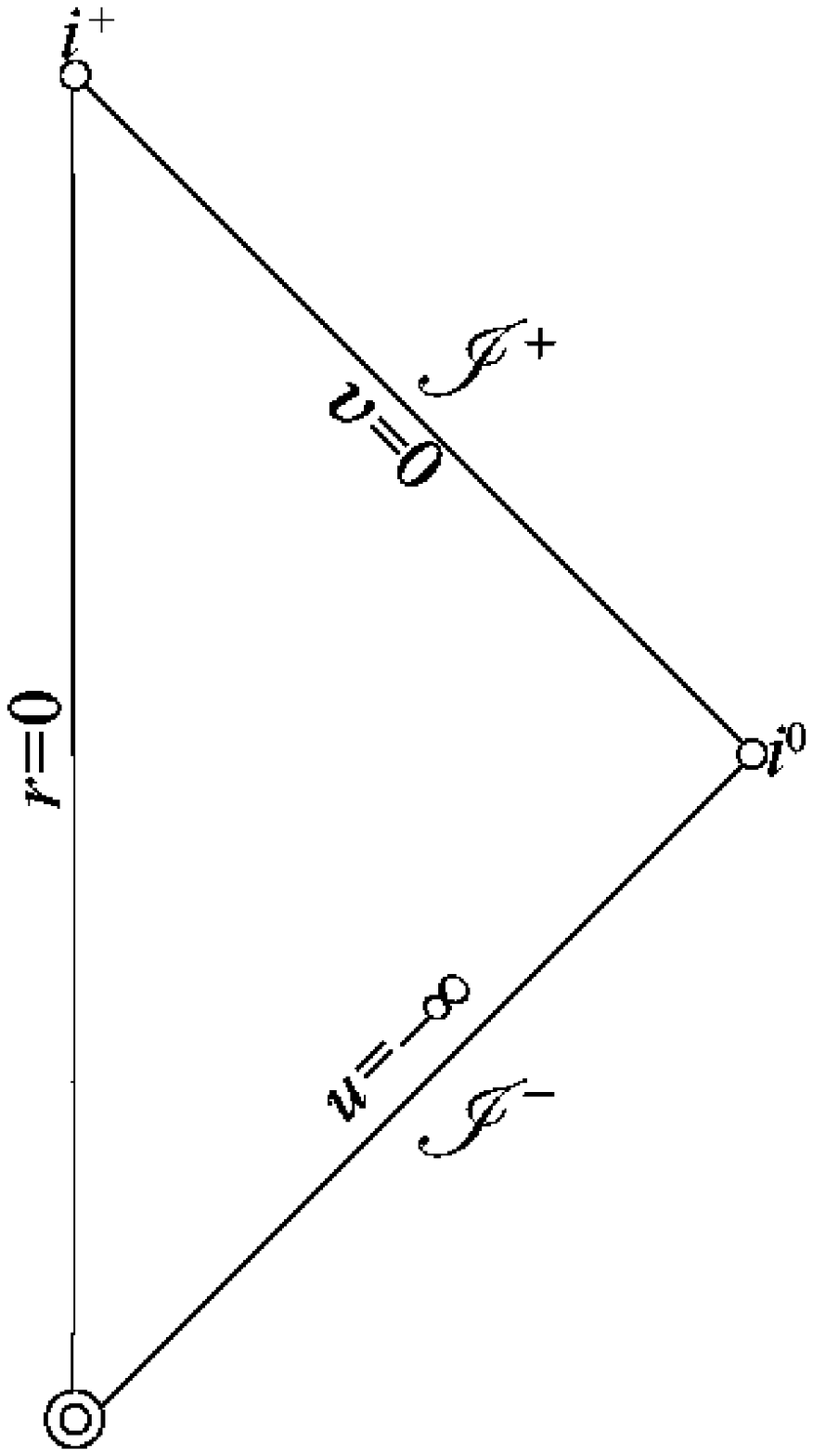}
\label{fg:penrose_kappa2gtr1o2_kappalt1}} &
\subfigure[]
{\includegraphics[width=0.25\textwidth]
{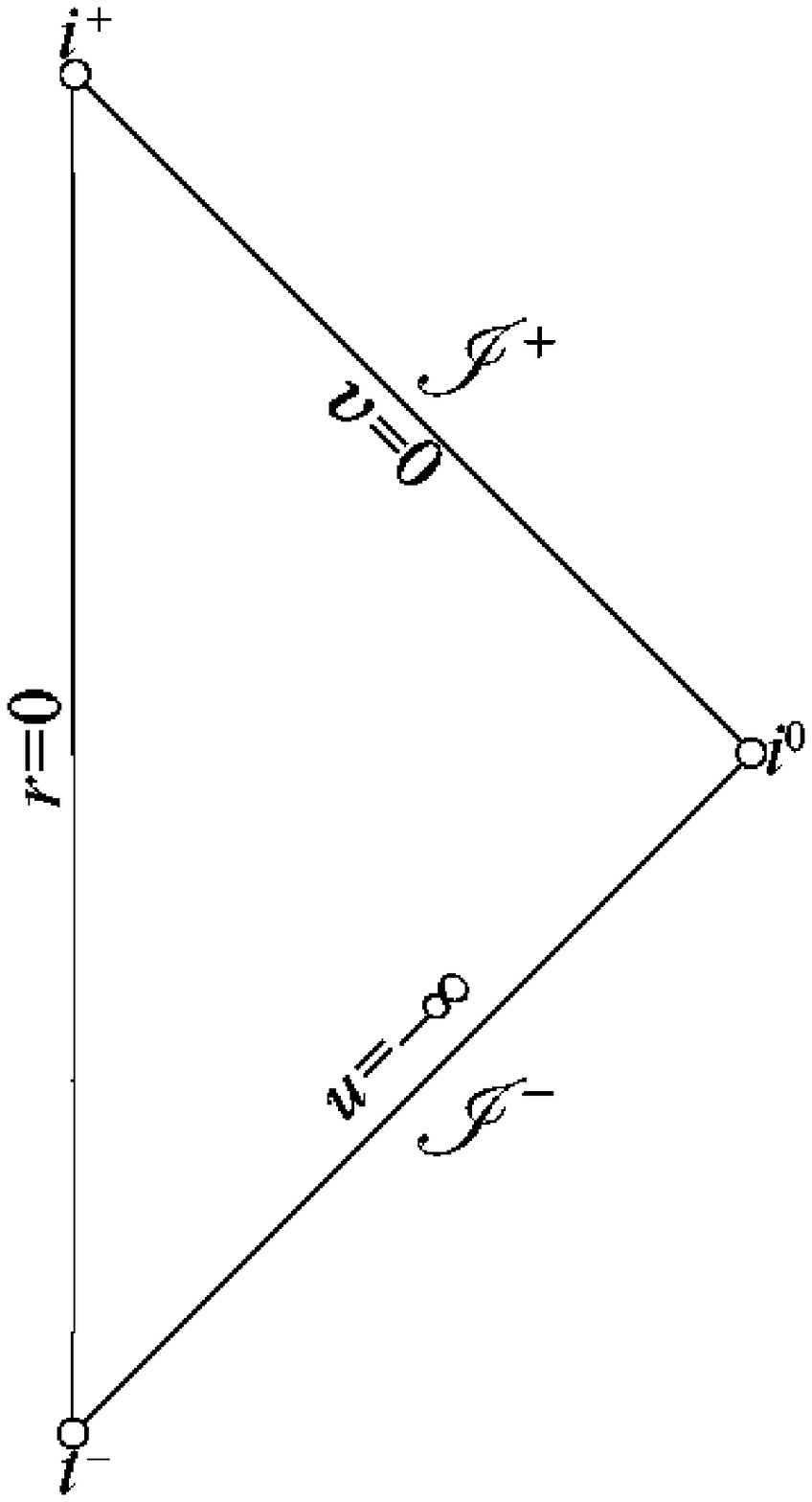}
\label{fg:penrose_kappa1}} &
\subfigure[]
{\includegraphics[width=0.25\textwidth]
{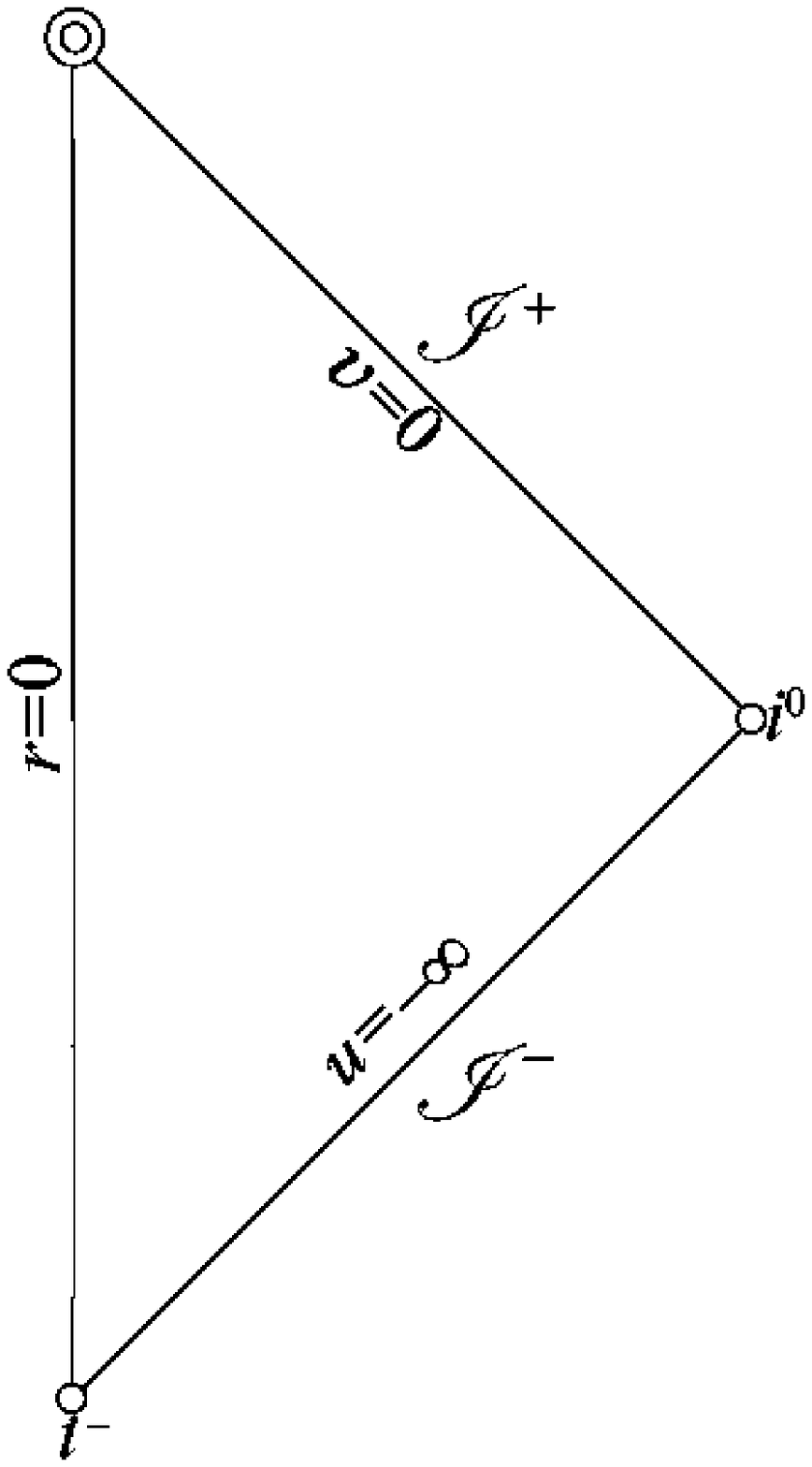}
\label{fg:penrose_kappagtr1}} & \\
\end{tabular}
\caption{The conformal diagrams of the two-parameter family of
solutions for different values of $\kappa$.
The conformal diagrams of the two-parameter family of solutions for (a) $0<\kappa^{2}<1/4$, $1/4<\kappa^{2}<3/8$, 
(b) $n:=1/[2(1-2\kappa^{2})]=2,3,4,\cdots$,
(c) $n>2$ but not an integer, 
(d) $\kappa\le -1/\sqrt{2}$, $1/\sqrt{2}\le \kappa<1$,
(e) $\kappa=1$,
and (f) $\kappa>1$.
The dashed lines denote cylindrical
trapping horizons.
The shaded and unshaded regions denote trapped and untrapped regions,
respectively.
The single circles denote timelike and spacelike infinities,
while the double circles and double lines denote spacetime singularities.
The solutions reduce to the Minkowski,
the cylindrical Milne and the Kasner solutions for $\kappa=0,1/2$
and $-1/2$, respectively. Choosing $\lambda\ne 0$ simply
introduces a conically singular axis.
For $3/8\le \kappa^{2}<1/2$ or $n\ge 2$,
if $n$ is an odd integer, there appears spacelike
curvature singularity with $r=0$ [see (b)]; if $n$ is an even integer,
it is replaced by noncurvature but quasiregular singularity [see (b)];
if $n$ is not an integer, then the spacetime
admits not analytic but only a $C^{[n]}$
extension beyond the null surface $u=0$ [see (c)],
where $[n]$ is the largest integer which
is no greater than $n$.}
\end{center}
\end{figure}

Also in this case, we introduce
$u$ and $v$ as
\begin{equation}
U=(-u)^{n}~~~~~~{\rm and}~~~~~~V=(-v)^{n},
\end{equation}
where $n= 1/[2(1-2\kappa^{2})]<0$ and we put the negative signs
to keep the increase of $u$ and $v$ corresponding to that of $U$ and $V$,
respectively.
Then, the line element is given by exactly the same form as
Eq.~(\ref{eq:3rd-metric}). The original domain is mapped to
$-\infty <u\le v<0 $. Since both $u=-\infty$ and $v=0$ correspond to
null infinities, the original domain describes the whole spacetime.
Noting $T=[(-v)^{n}+(-u)^{n}]/2$ and $X=[(-v)^{n}-(-u)^{n}]/2$, we find that
$T=X=0$ corresponds to $v=u=-\infty$, while
$(T,X)=(\infty,0)$ corresponds to $v=u=0$.
The conformal diagrams of the solutions for different values of $\kappa$
are therefore given in Figs.~\ref{fg:penrose_kappa2gtr1o2_kappalt1},
\ref{fg:penrose_kappa1} and \ref{fg:penrose_kappagtr1}.
There is no cylindrical trapping horizon and
the whole spacetime is untrapped in this case.

\vspace{1cm}
In summary of this section, the two-parameter family of solutions describe
a variety of global structures. They are classified in terms of $\kappa$,
or equivalently, $n=[2(1-2\kappa^{2})]^{-1}$.
For $\kappa=0$, $1/2$ and $-1/2$, the solution reduces to the Minkowski,
the cylindrical Milne and the Kasner solutions, respectively.
If $0<\kappa^{2}<1/4$ or $1/4<\kappa^{2}<3/8$,
the spacetime describes the interior of the exploding (imploding)
cylindrical shell of gravitational waves.
For $3/8\le \kappa^{2}<1/2$, we have the following three cases:
if $n=3,5,7,\cdots$,
the spacetime describes the collapsing gravitational waves
to a spacelike singularity or exploding gravitational waves
from a spacelike curvature singularity;
if $n=2,4,6,\cdots$, the spacetime structure is quite similar
to the odd $n$ case but the spacelike curvature singularity
is replaced by a quasiregular one;
if $n>2$ is not an integer,
the spacetime does not admit an analytic extension beyond
the null surface.
For $\kappa^{2}\ge 1/2$ and $\kappa\ne 1$, 
the conformal diagram is similar to the Minkowski
one except for that a singularity replaces a timelike infinity.
For $\kappa=1$, the conformal diagram is similar to the Minkowski
one and the whole spacetime is regular.

It is interesting to note that
the present analysis proceeds quite analogously to Wang's~\cite{Wang:2003}
for cylindrically symmetric
self-similar solutions with a massless scalar field
in (3+1)-dimensions
and hence Hirschmann's et al.~\cite{Hirschmann:2004}
for circularly symmetric
self-similar solutions with a massless scalar field
in (2+1)-dimensions
although the system and the result are both different in detail.
This follows from the fact that the governing system of partial
differential equations are quite similar for these
systems.
It is also interesting to note that the present solution generically
involves singularities, because one does not 
usually get a singularity from the collapse of 
cylindrical waves (global regularity of 
Einstein-Rosen waves: see e.g. Ashtekar 
et al.~\cite{Ashtekar:1997}).
In order to generate the singularity, initial data on the Cauchy surface 
must be non asymptotically flat. So we have a sort of converse result: 
if we allow non asymptotically flat initial data, then a 
singularity can form solely from the collapse of gravitational waves.
Furthermore, if trapped surfaces are initially absent, then they 
cannot form (see Thorne~\cite{Thorne:1965}). 
In the relevant cases (Fig.~\ref{fg:penrose_integern}, integer $n\ge
2$), we have a nonsingular Cauchy surface which already contains trapped cylinders.

\section{Conclusions}

In this paper, we have studied self-similar
vacuum spacetimes in whole-cylinder symmetry:
self-similar Einstein-Rosen waves. The primary motivation was to determine possible (intermediate) asymptotic endstates for more general Einstein-Rosen waves, and for other cylindrical systems. That is, we wish to study the self-similarity hypothesis in cylindrical symmetry, especially in the context of gravitational collapse. There is a considerable body of evidence for the hypothesis in spherical collapse (see e.g.~\cite{Carr:2005}) and there is also evidence for the hypothesis in the context of cosmological models: 
Hewitt et al. have shown that among a class of cylindrical inhomogeneous cosmological models, there are self-similar models which are asymptotic endstates for the general class~\cite{Hewitt:1991}.

Assuming a homothetic vector orthogonal to the
cylinders of symmetry,
we have obtained the standard form of the metric in cylindrically
symmetric self-similar spacetimes. We have then applied this
to the vacuum case and obtained solutions.
In fact, the obtained solutions are all flat.
We have explicitly shown that the spacetimes are part of
the Minkowski spacetime with a regular or conically singular axis
and with trivial or nontrivial topology. Although such spacetimes 
can emerge as the endstate of complete cylindrical gravitational collapse - for example, of cylindrical null dust~\cite{Nolan:2002}, this cannot be interpreted as evidence for the self-similarity hypothesis as the spacetimes are flat and therefore only trivially self-similar.

Using the obtained self-similar expression for part of the Minkowski
spacetime, we have argued that the $C$ energy which is supposed to
represent the gravitational energy per specific length
of the cylindrically symmetric
spacetime is subject to the choice of the translational Killing
vector even if one chooses the same regular axis. We have also
discussed that there exists a cylindrical trapping horizon in the
Minkowski spacetime and that the notion of trapping horizons might
not be useful for defining black holes in cylindrically symmetric
spacetimes -- at least, not in the case where the marginal two-surfaces
foliating the horizon are cylinders.

Next, we have extended the analysis to the more general
class of Einstein-Rosen waves,
still respecting some kind of scaling behavior.
Assuming a regular or conically singular axis,
we have obtained a two-parameter family of non flat
self-similar solutions, where the homothetic vector
is not orthogonal to the
cylinders
in general.
We have seen that the solution physically describes
the interior of
the exploding (imploding) shell of gravitational waves or
the collapse (explosion) of gravitational
waves depending on the parameter choice.
Additionally, as a special case we have obtained a solution with a
non azimuthal and non translational Killing vector
which is not orthogonal to the 
cylinders.
There is also a discrete subset of solutions which exhibit
the collapse of gravitational waves
developed from nonsingular initial data on a spacelike Cauchy surface.

We conclude that self-similar Einstein-Rosen waves can describe nontrivial dynamics of gravitational waves if and only if the homothetic vector is not orthogonal to the cylinders of symmetry. Although the original proposal for the self-similarity hypothesis in general relativity is restricted in spherical symmetry, it is also likely that some of these self-similar solutions can describe the asymptotic behavior of more 
general solutions even in cylindrical symmetry. In fact, recent
numerical simulations~\cite{Nakao:2009} strongly suggest that the
asymptotic behavior of a dispersing gravitational wave within the null hypersurface $t^2 = x^2$ after the collapse of a dust cylinder 
is well approximated by a member of the family of solutions obtained here with 
$\kappa = -0.0206$ and $\lambda \ne 0$ (see Figs. 8 and 9
of~\cite{Nakao:2009}). The present paper clarifies 
that this asymptotic solution belongs to the family of self-similar
Einstein-Rosen 
waves with a non cylindrical homothetic vector and a conical singularity
and that this 
asymptotic solution corresponds to gravitational waves inside the
exploding shell
of gravitational waves as shown in Fig. 3(a). 
Hence, it would be reasonable to 
generalize the self-similarity hypothesis -- including in the context of
gravitational collapse -- as follows: under certain physical
circumstances, solutions will naturally 
evolve to a self-similar form not only 
in spherical symmetry but also in a variety of 
symmetry classes.

\acknowledgments

The authors would like to thank M.~Narita and H.~Maeda for fruitful
discussions. The authors would also like to thank the anonymous referee
for helpful comments.
TH was partly supported by the Grant-in-Aid for Scientific
Research Fund of the Ministry of Education, Culture, Sports, Science
and Technology, Japan [Young Scientists (B)
18740144 and 21740190]. BN gratefully
acknowledges support from the Rikkyo University Research
Associate Program and the Department of Physics, Osaka City University.

\appendix

\section{Timelike geodesics with infinite spatial acceleration
in the plane symmetric closed Milne universe}

\begin{figure}[htbp]
\begin{center}
\includegraphics[width=0.4\textwidth]{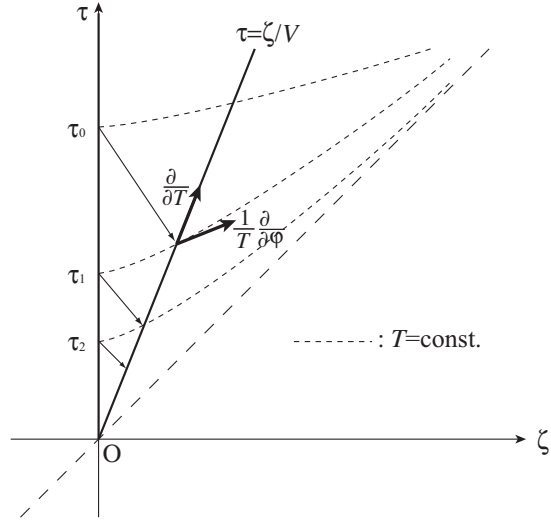}
\caption{\label{fg:accelerated_geodesic}
The closed Milne universe in the Minkowski spacetime.
}
\end{center}
\end{figure}

The line element of the two-dimensional closed Milne universe is given by
\begin{equation}
ds^2=-dT^2+T^2d\varphi^2,
\end{equation}
where $0\leq\varphi<2\pi$ and $\varphi=0$ and $\varphi=2\pi$ with the same $T$ are
identified with each other. This is obtained from Eq.~(\ref{eq:nontrivial_Minkowski_1}) or (\ref{eq:nontrivial_Minkowski_2}) by omitting the two-dimensional
plane part.
The Milne universe is locally identical to the Minkowski spacetime.
In fact, the coordinate transformation
\begin{equation}
\tau=T\cosh\varphi~~~~~{\rm and}~~~~~\zeta=T\sinh\varphi \label{eq:coordtrans}
\end{equation}
leads to the line element of standard form of the Minkowski spacetime
\begin{equation}
ds^2=-d\tau^2+d\zeta^2
\end{equation}
In this $(\tau,\zeta)$-coordinate system, the curves $\zeta=0$ and
$\tau=\zeta/V$ are identified with each other, where
$V=\tanh 2\pi$. More precisely, $(\tau,0)$ is identified with
$(\tau/\sqrt{1-V^{2}}, \tau V/\sqrt{1-V^{2}})$, where
and hereafter we use the standard Cartesian
coordinates $(\tau,\zeta)$ of the Minkowski spacetime.
See Fig.~\ref{fg:accelerated_geodesic}.

We consider a past-directed timelike geodesic which
starts from  $p_0:(\tau,\zeta)=(\tau_0,0)$ with a
unit tangent vector
\begin{equation}
{\bf u}={\bf u}_{(0)}:= -\frac{1}{\sqrt{1-v_0^2}}
\frac{\partial}{\partial \tau}+
\frac{v_0}{\sqrt{1-v^2_0}}\frac{\partial}{\partial \zeta},
\end{equation}
where $0<v_0<1$ is assumed.
This geodesic and the curve $\tau=\zeta/V$ intersect at the event
\begin{equation}
\tilde{p}_0:(\tau,\zeta)
=\left(\frac{v_0\tau_0}{V+v_0},\frac{Vv_0\tau_0}{V+v_0}\right).
\end{equation}
This event is identified with
\begin{equation}
p_1:(\tau,\zeta)=\left(\tau_1,0\right),
\end{equation}
where
\begin{equation}
\tau_1:=\frac{v_0\sqrt{1-v_0^2}}{V+v_0}~\tau_0~.
\end{equation}

We can determine the tangent to the geodesic at $p_1$ as follows. In coordinates $(T,\varphi)$, the points $\tilde{p}_0:(T_0,0)$ and $p_1:(T_0,2\pi)$ are identified. We must also identify the unit vector fields
\begin{equation} \left.\frac{\partial}{\partial T}
\right|_{\tilde{p}_0}= \left.\frac{\partial}
{\partial T}\right|_{p_1},\quad
\left.\frac{1}{T}\frac{\partial}{\partial \varphi}\right|_{\tilde{p}_0}
= \left.\frac{1}{T}\frac{\partial}{\partial \varphi}\right|_{p_1}.\label{eq:idvfs}\end{equation}

The coordinate transformation (\ref{eq:coordtrans}) leads to
\begin{eqnarray} \frac{\partial}{\partial T}= \frac{\tau}{T}
\frac{\partial}{\partial \tau} +\frac{\zeta}{T}
\frac{\partial}{\partial \zeta},\qquad
\frac{1}{T}\frac{\partial}{\partial \varphi}&=& \zeta
\frac{\partial}{\partial \tau} +\tau\frac{\partial}{\partial \zeta},
\label{eq:vfstrans}
\end{eqnarray}
and the corresponding inverse relationship:
\begin{equation}
\frac{\partial}{\partial \tau} =\frac{\tau}{T}\frac{\partial}{\partial T}
-\frac{\zeta}{T^3}\frac{\partial}{\partial \varphi},\qquad
\frac{\partial}{\partial \zeta} =-\frac{\zeta}{T}
\frac{\partial}{\partial T}+\frac{\tau}{T^3}\frac{\partial}{\partial \varphi}.
\label{eq:vfsinvtrans}
\end{equation}

We can use this last equation to expand
$\left.u^\mu_{(0)}\right|_{\tilde{p}_0}$ in terms of
$\frac{\partial}{\partial T}$ and
$T^{-1}\frac{\partial}{\partial \varphi}$,
invoke the identification (\ref{eq:idvfs}), revert to $(\tau,\zeta)$
coordinates by using (\ref{eq:vfstrans}) and
hence obtain the unit tangent to the geodesic at $p_1$ in
coordinates $(\tau,\zeta)$ as
\begin{equation}
{\bf u}_{(1)}=-\frac{1}{\sqrt{1-v_1^2}}
\frac{\partial}{\partial \tau}+
\frac{v_1}{\sqrt{1-v^2_1}}
\frac{\partial}{\partial \zeta}
\end{equation}
where
\begin{equation}
v_1=\frac{v_0+V}{1+Vv_0}.
\end{equation}

It is also straightforward to show that in the transition
from $(\tau_0,0)$ to $(\tau_1,0)$, a proper time of duration
\[ s_0: = \frac{V\sqrt{1-v_0^2}}{v_0+V}\tau_0\]
elapses along the geodesic.

Thus, this timelike geodesic goes through the points
$(\tau,\zeta)=(\tau_1,0),~(\tau_2,0),\cdots$.
We can derive the recursion relations
\begin{eqnarray}
\tau_{i+1}&=&\frac{v_0\sqrt{1-v_i^2}}{V+v_i}~\tau_i, \label{eq:t-rec}\\
v_{i+1}&=&\frac{v_i+V}{1+Vv_i},\label{eq:v-rec}\\
s_{i+1}&=&\frac{v_i\sqrt{1-v_i^2}\sqrt{1-V^2}}{v_i+2V+v_iV^2}s_i.\label{eq:tau-rec}
\end{eqnarray}

Here we introduce a new variable
\begin{equation}
\delta_i:=1-v_i
\end{equation}
and rewrite Eqs. (\ref{eq:t-rec}) and (\ref{eq:v-rec}) as
\begin{eqnarray}
\tau_{i+1}&=&\frac{(1-\delta_i)
\sqrt{\delta_{i}(2-\delta_{i})}}{V+1-\delta_{i}}~\tau_i, \label{eq:t-rec_2}\\
\delta_{i+1}&=&\frac{1-V}{1+V(1-\delta_i)}~ \delta_i. \label{eq:delta-rec_2}
\end{eqnarray}
 From Eq.~(\ref{eq:delta-rec_2}), we have
\begin{equation}
\frac{\delta_{i+1}}{\delta_i}=\frac{1-V}{1+V(1-\delta_{i})}.
\end{equation}
Since $0<\delta_{0}<1$ and $0<V<1$, we have $0<\delta_{1}<\delta_{0}$
and hence $0<\delta_{i+1}<\delta_{i}$. This implies
\begin{equation}
0<\delta_{i+1}<\left[\frac{1-V}{1+V(1-\delta_0)}\right]^{i}\delta_{0}.
\end{equation}
Therefore,
\begin{equation}
\lim_{i\to\infty}\delta_{i}=0.
\end{equation}
As for $\tau_{i}$, since $0<\delta_{i}< \delta_{0}<1$ for $i\ge 1$,
we have from Eq.~(\ref{eq:t-rec_2})
\begin{equation}
0<\frac{\tau_{i+1}}{\tau_{i}}=\frac{(1-\delta_i)\sqrt{\delta_{i}(2-\delta_{i})}}{V+1-\delta_{i}}
<\frac{1-\delta_i}{V+1-\delta_{i}}<1.
\end{equation}
Therefore
\begin{equation}
\lim_{i\rightarrow\infty}\tau_i=0.
\end{equation}

Thus the three-velocity $v$ of the timelike geodesic
which approaches $p_\infty:(\tau,\zeta)=(0,0)$ becomes
asymptotically the speed of light. The spatial part of the past-directed
timelike geodesic is infinitely
accelerated in an approach to the origin $p_\infty:(\tau,\zeta)=(0,0)$.

We note however that $p_\infty$ lies at a finite time in the past along the history of the geodesic. The total proper time that elapses along the geodesic is given by
\[ s = \sum_{i=0}^\infty s_i.\]
However, \[\frac{s_{i+1}}{s_i} =
\frac{v_i\sqrt{1-v_i^2}\sqrt{1-V^2}}
{v_i+2V+v_iV^2}\to 0,\quad i\to\infty,\]
so the series converges: $s <+\infty$.

Thus, there is no unique
extension of the geodesics beyond $(\tau,\zeta)=(0,0)$ 
and this behavior of geodesics 
is quite analogous to that around a conical singularity. This is solely due to the 
topological identification in $\varphi$ and not related 
to the blow up of curvature. This corresponds to a  quasiregular singularity defined by
Ellis and Schmidt~\cite{Ellis:1977}.

\newpage

\section*{Erratum}

We made a typographical error in the first term on the right hand side
of Eq.~(5.11). This is rectified as follows:
\[
ds^2=-[2(1-2\kappa^{2})]^{-2}
e^{2\lambda}\left(\frac{V+U}{2}\right)^{4\kappa(2\kappa-1)}
dudv
+\left(\frac{V+U}{2}\right)^{2(1-2\kappa)}
\left(\frac{V-U}{2}\right)^{2}
d\varphi^2
+\left(\frac{V+U}{2}\right)^{4\kappa}dz^2.
\]

Also in the caption of Fig. 3, we erroneouly wrote that Fig. 3(b) 
showed the conformal diagram for $n=2, 3, 4, \cdots$. 
In reality, Fig. 3(b) shows the conformal diagram only 
for $n=3,5,7,\cdots$. As for $n=2,4,6,\cdots$,
the conformal diagram is given by the following figure, where 
unshaded regions denote untrapped regions.
For this case, the extended region $-v<u<0$ is 
untrapped as is the original region $0<u\le v$. 
The null surface $u=0$ is a trapping horizon.
The timelike surface $r=0$ is a regular or conically singular axis, 
while the spacelike surface $r=0$ is noncurvature quasiregular 
singularity.
\begin{figure}[htbp]
\begin{center}
\includegraphics[width=0.4\textwidth]
{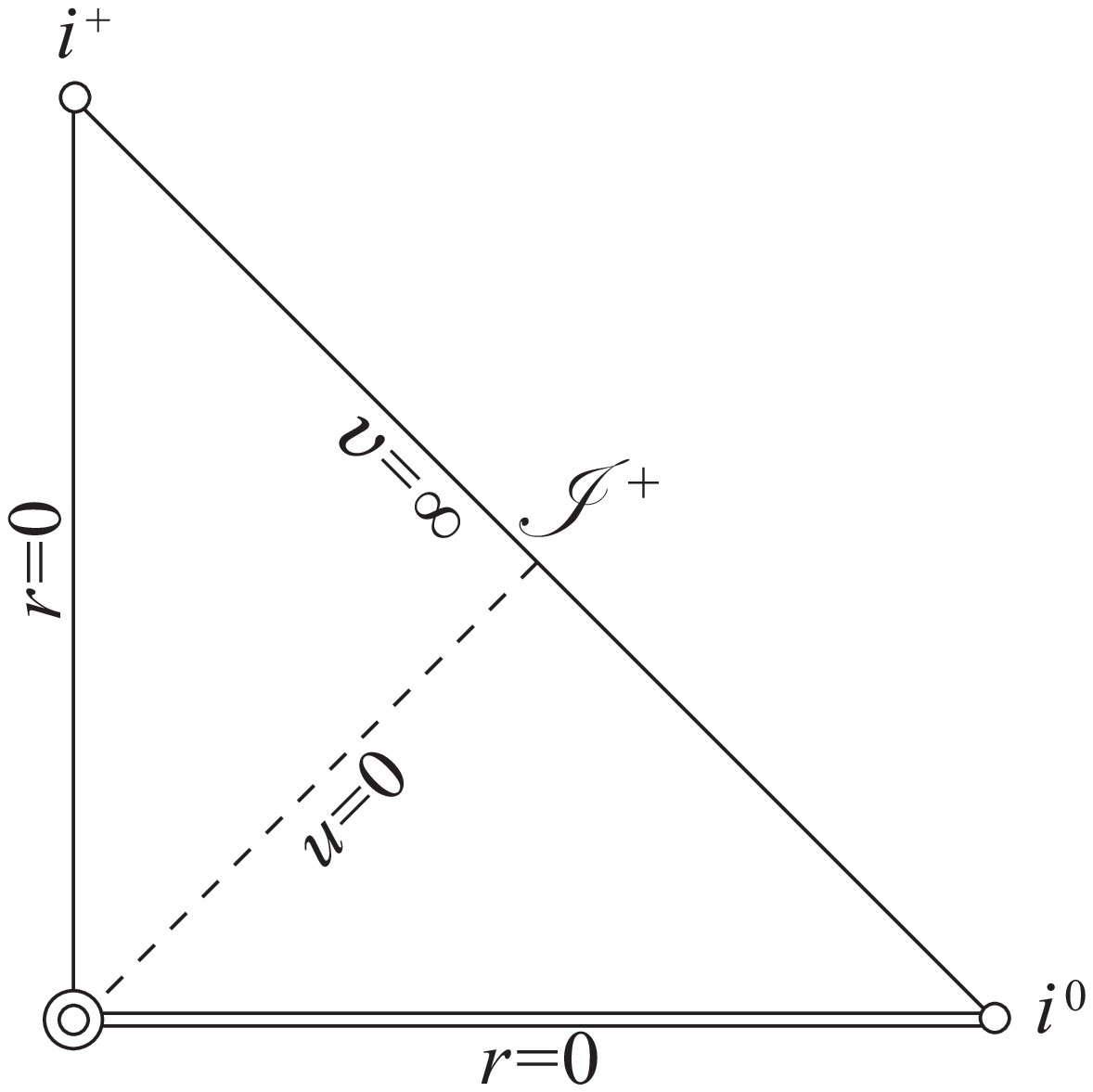}
\end{center}
\end{figure}


\begin{thebibliography}{99}

\bibitem{Einstein:1937}
A.~Einstein and N.~Rosen,
J. Franklin Inst. {\bf 223}, 43 (1937).

\bibitem{Kennefick:2007}
D.~Kennefick, {\it Travelling at the Speed of Thought. Einstein and
the Quest for Gravitational Waves}. (Princeton University Press, Princeton, 2007).

\bibitem{Thorne:1965}
K.~S.~Thorne, Phys. Rev. {\bf 138}, B251 (1965).

\bibitem{Melvin:1965}
M.~A.~Melvin, Phys. Rev. {\bf 139}, B225 (1965).

\bibitem{Apos:1992}
T.~A.~Apostolatos and K.~S.~Thorne, Phys. Rev. D{\bf 46}, 2435
(1992).

\bibitem{echeverria:1993}
F.~Echeverria, Phys. Rev. D{\bf 47}, 2271 (1993).

\bibitem{letelier:1994}
P.~S.~Letelier and A.~Wang, Phys. Rev. D{\bf 49}, 5105 (1994).

\bibitem{Chiba:1996}
T.~Chiba, Prog.~Theor.~Phys. {\bf 95}, 321 (1996).

\bibitem{Ashtekar:1997}
A.~Ashtekar, J.~Bicak and B.~G.~Schmidt, 
Phys.~Rev.~D{\bf 55}, 687 (1997).

\bibitem{Hayward:2000}
S.~A.~Hayward, Classical Quantum Gravity {\bf 17}, 1749 (2000).

\bibitem{Nolan:2002} B.~C.~Nolan, Phys. Rev. D{\bf 65} 104006 (2002).

\bibitem{Wang:2003}
A.~Wang, Phys.~Rev. D{\bf 68} 064006 (2003).

\bibitem{Nakao:2004} K.~Nakao and Y.~Morisawa, Classical Quantum Gravity {\bf 21},
2101 (2004).

\bibitem{Nakao:2005} K.~Nakao and Y.~Morisawa, Phys. Rev. D{\bf 71} 124007
(2005).

\bibitem{Kurita:2006} Y.~Kurita and K.~Nakao, Phys. Rev. D{\bf 73}
064022 (2006).

\bibitem{Nakao:2007} K.~Nakao, Y.~Kurita, Y.~Morisawa and T.~Harada,
Prog. Theor. Phys. {\bf 117}, 75 (2007).

\bibitem{Nakao:2008}K.~Nakao, D.~Ida and Y.~Kurita, Phys. Rev. D{\bf 77}
044021 (2008).

\bibitem{Carr:1993}
B.~J.~Carr, (unpublished).

\bibitem{Carr:2005}
B.~J.~Carr and A.~A.~Coley, Gen. Relativ. Grav. {\bf 37}, 2165 (2005).

\bibitem{Harada:2001} T.~Harada and H.~Maeda, Phys. Rev. D{\bf 63} 084022
(2001).

\bibitem{Snajdr:2006} M. Snajdr, Classical Quantum Gravity {\bf 23}, 3333 (2006).

\bibitem{Gundlach:2007} C.~Gundlach and J.~M.~Mart\'in-Garc\'ia,
Living Rev. Rel.
{\bf 10},  5 (2007). URL (cited on 4th September 2008):
http://www.livingreviews.org/lrr-2007-5

\bibitem{Wainwright:1997} J.~Wainwright and G.~F.~R.~Ellis {\it
Dynamical Systems in Cosmology}, (Cambridge University Press, 1997)

\bibitem{Sharif:2005a}
M.~Sharif and S. Aziz, Int. J. Mod. Phys. A{\bf 20}, 7579 (2005).

\bibitem{Sharif:2005b}
M.~Sharif and S. Aziz, Int. J. Mod. Phys. D{\bf 14}, 1527 (2005).

\bibitem{Nolan:2007}
L.~V.~Nolan, {\em Cylindrically Symmetric Models of Gravitational
Collapse}, PhD thesis, Dublin City University, 2007 (unpublished).

\bibitem{Nakao:2009}
K.~Nakao, T.~Harada, Y.~Kurita and Y.~Morisawa, arXiv:0905.3968 [Prog. Theor. Phys. (to be published)].

\bibitem{Stephani:2003}
H.~Stephani, D.~Kramer, M.~A.~H.~MacCallum, C.~Hoenselaers and
E.~Herlt, {\it Exact solutions of Einstein's field equations},
(Cambridge University Press, 2003).

\bibitem{Senovilla:1997}
J.~M.~M.~Senovilla, Gen. Relativ. Grav. {\bf 30}, 701 (1998).

\bibitem{Ellis:1977}
G.~F.~R.~Ellis and B.~G.~Schmidt, 
Gen. Relativ. Grav. {\bf 8}, 915 (1977).

\bibitem{Hirschmann:2004}
E.~W.~Hirschmann, A.~Wang and Y.~Wu,
Classical Quantum Gravity {\bf 21}, 1791 (2004).

\bibitem{Hewitt:1991}
C.~G.~Hewitt, J.~Wainwright and M.~Glaum, Classical Quantum Gravity {\bf 8}, 1505 (1991).


\end{thebibliography}
\end{document}